\documentclass[14pt]{article}
\usepackage[english]{babel}
\usepackage[utf8x]{inputenc}

\usepackage{latexsym}
\usepackage{amsfonts}
\usepackage{amsmath}
\usepackage{amssymb}

\usepackage{graphicx}
\usepackage[colorinlistoftodos]{todonotes}
\usepackage{dsfont}
\usepackage{ mathrsfs }
\usepackage[colorlinks=true, allcolors=blue]{hyperref}
\DeclareGraphicsExtensions{.pdf, .png, .jpg}
\newcommand{\E}{\mathrm{e}}
\newcommand{\I}{\mathrm{i}}

\newcommand{\be}{\begin{equation}}
\newcommand{\ee}{\end{equation}}
\newcommand{\bean}{\begin{eqnarray*}}
	\newcommand{\eean}{\end{eqnarray*}}

\newcommand{\bea}{\begin{eqnarray}}
\newcommand{\eea}{\end{eqnarray}}

\usepackage{indentfirst}

\usetikzlibrary{shapes.misc}

\tikzset{cross/.style={cross out, draw=black, minimum size=2*(#1-\pgflinewidth), inner sep=0pt, outer sep=0pt},
cross/.default={1pt}}

\usepackage[left=2.7cm, top=3cm, right=2.7cm]{geometry}
\title{Ternary and Binary Representation of Coordinate and Momentum in Quantum Mechanics}
\author{M. G. Ivanov\thanks{\href{mailto:ivanov.mg@mipt.ru}{ivanov.mg@mipt.ru}}, A. Yu. Polushkin\thanks{\href{mailto:polushkin.ayu@phystech.edu}{polushkin.ayu@phystech.edu}}\\
\small Moscow Institute of Physics and Technology}

\begin{document}

\maketitle

\begin{abstract}
To simulate a quantum system with continuous degrees of freedom on a quantum computer based on
quantum digits, it is necessary to reduce continuous observables (primarily coordinates and momenta) to discrete
observables. We consider this problem based on expanding quantum observables in series in powers of
two and three analogous to the binary and ternary representations of real numbers. The coefficients of the series (``digits'') are, therefore, Hermitian operators. We investigate the corresponding quantum mechanical operators and the
relations between them and show that the binary and ternary expansions of quantum observables automatically leads
to renormalization of some divergent integrals and series (giving them finite values).
\end{abstract}

\section{Introduction}

% пока передрал с двоичной статьи
When future applications of quantum computers are discussed, attention is often focused primarily
on cryptanalysis tasks. Nevertheless, by the time quantum computers become sufficiently powerful, \textit{post-quantum cryptography}, which is resistant to cryptanalysis on a quantum computer, will be universally
introduced \cite{post-q}. In this regard, we can expect that quantum computers’ primary use will turn out to be “peaceful” (unrelated to the breaking of ciphers). In particular, much attention will be paid to modeling quantum systems in accordance with Feynman’s original idea \cite{Feynman}.

Modeling quantum systems is a very relevant task from the standpoint of practical applications such
as quantum chemistry, the creation of new materials, quantum biophysics, the development of new drugs,
nuclear physics, and elementary particle physics. Many of these applications require discretizing continuous
quantum observables, coordinates, and momenta, which should be described in a quantum computer using a
set of discrete quantum cells (qubits and/or qutrits and other qudits). Apparently, momentum operators are
differential operators, and we can use well-known difference schemes from computational mathematics to
discretize them (see, e.g., \cite{simulating_qm_at_qc}, \cite{elementary_qm_at_lattice}). 
But such a direct approach disregards the specifics of quantum mechanics,
in which the momentum operator is a generator of shifts along the corresponding coordinate.

Here, we construct representations of continuous quantum observables based on an expansion in powers of two and three, analogous to the binary and ternary representations of real numbers. 
An individual binary or ternary digit of the observable expansion is itself observable and described by a single quantum bit or trit. 
We construct the operators of binary and ternary digits of the coordinate and momentum on a lattice and on the line in an explicit form, for which we obtain commutation relations. 
We construct a binary and ternary integral representation of real numbers and quantum observables. 
In the representation of observables in the form of binary and ternary series and integrals, we
naturally assign finite values to some formally divergent expressions, i.e., we introduce a renormalization.
This renormalization method applies to finite quantities (which are related to the ambiguity in choosing the space $\mathbb{Z}_N$), allowing for calculating the lattice's renormalized quantities.

% пояснения о самоцитировании
This paper is a further development of the ideas introduced in \cite{bin-ivanov}, which considered the binary decomposition of coordinates and momentum in quantum mechanics. Here, we investigate the ternary decomposition, and also consider, in more detail,  some properties of the binary decomposition. For the convenience of comparing old and new results, we follow the order of presentation as close as possible to the paper \cite{bin-ivanov}. So the paper is full of self-citation, which is not further specified.

The generalization from binary representation to ternary was very straightforward, but at the same time it allowed us to better understand the overall picture, including the binary case.

%(Надо дописать)
\section{Coordinates and momenta on a finite lattice}
Here and hereafter, we use the coordinate representation (unless otherwise stated) and assume $h = 1$ and $\hbar = 1/2\pi$, $1/\hbar = 2\pi$ for the Planck constant.
\subsection{Coordinates and momenta on a finite lattice}
We assume that the coordinate is described by $n$ digit-qutrit, and the coordinate lattice consists of $N = 3^n$ nodes, which we assume to be cyclic (after the last one comes the first). If we suppose that the \textit{coordinate lattice constant} is $\Delta x = 3^{-n_{-}}$, then the \textit{lattice period} is equal to $\Xi = N\Delta x = 3^{n_{+}}$, $n_{+} = n - n_{-}$. We assume that the values $x$ range from 0 to $\Delta x \cdot (N - 1)$.

On the coordinate lattice, a natural addition operation is induced from $\mathbb{Z}_{N}$, for which $x = x + \Xi$. It is possible to use other representations of the lattice $\Delta x \cdot \mathbb{Z}_{N}$ by real numbers. For example, taking the equivalence of $x$ and $x + \Xi$ into account, we further need a representation in which $x$ ranges in $\{-3^{n-1}\Delta x,\dots,-\Delta x,0,\Delta x,2\Delta x,\dots,(2 \cdot 3^{n-1}-1)\Delta x\}$. The power series for the coordinate on a finite lattice is finite:
\begin{equation}
    x = \sum\limits_{s = -n_{-}}^{n_{+} - 1}x_{s}3^{s} = \sum\limits_{s = -n_{-}}^{n_{+} - 1}\mathbf{t}(s,x)3^{s}.
\end{equation}
Here, $x_{s} = \mathbf{t}(s,x)$ is the $s$-th digit in the {\bf t}ernary expansion of $x$. We sometimes specify a range of powers of three that defines a lattice and write $x_{s} = \mathbf{t}_{n_{-}n_{+}}(s,x)$.
 
We introduce the coordinate basis $\{|x\rangle \}_{x \in \Delta x \cdot \mathbb{Z}_{N}}$ for the functions defined on the lattice:
\begin{equation}
    \hat{x}|x\rangle = x|x\rangle, \quad \langle x^{\prime}|x^{\prime \prime} \rangle = \delta_{x^{\prime},x^{\prime \prime}}, \quad \psi(x) = \langle x|\psi \rangle, \quad x \in \Delta x \cdot \mathbb{Z}_{N}.
\end{equation}
We represent wave functions (ket vectors) in the forms of columns whose rows are ordered in decreasing order of $x$. Thus, if $x$ varies from $0$ to $(N - 1)\Delta x$, then
\begin{equation}\psi (x) =\left(
    \begin{array}{c}
           \psi((N - 1)\Delta x)\\
           \psi((N - 2)\Delta x)\\
           .\\
           .\\
           .\\
            \psi(\Delta x)\\
            \psi(0)\\
    \end{array}\right).
\end{equation}

\subsection{Momentum lattice}
We define the momentum operator $\hat{p}$ as the generator of the shifts $\widehat{T}_{A}$ along the coordinate lattice:
\begin{equation}
    \widehat{T}_{A} \psi(x) = \psi(x + A), \qquad \widehat{T}_{A} =\E^{2\pi \I A \hat{p}}, \quad A \in \Delta x \cdot \mathbb{Z}. 
\end{equation}
Such operators were considered in Weyl's classic book \cite{Weyl} and more detailed later by Schwinger \cite{Schwinger}.

Because the coordinate lattice is periodic, the shift by the period $\Xi$ must be identity transformation, i.e., for eigenvalues of operator $\hat{p}$, we have 
$\Xi\cdot p\in\mathbb{Z}$.
This gives the \textit{momentum step} $\Delta p$,
\begin{equation}
    \Xi \cdot \Delta p = 1, \quad \Delta p = 3^{-n_{+}}, \quad \Delta p \cdot \Delta x = \frac{1}{N} = 3^{-n}
\end{equation}
The number of points in the spectrum of momentum is the same as for the coordinate, i.e., for momentum,
we have a poriodic lattice with the same number of nodes but a different period $\Pi = \Delta p \cdot N = 3^{n_{-}}$, $\Pi \Xi = N$. 
The momentum lattice is denoted by $\Delta p \cdot \mathbb{Z}_{N}$. 
The power series for the momentum is also finite:
\begin{equation}
    p = \sum\limits_{r = -n_{+}}^{n_{-} -1}p_{r}3^{r} = \sum\limits_{r = -n_{+}}^{n_{-} -1} \mathbf{t}(r,p)3^r.
\end{equation}
Here, $p_r=\mathbf{t}(r,p)$ is the $r$-th digit in the \textbf{t}ernary expansion of $p$.
We sometimes specify a range of powers of three, which defines the lattice,  and write $p_r = \mathbf{t}_{n_+n_-}(r,p)$.

\subsection{Minimum shift}
The minimum shift $\widehat{T}_{\Delta x}$ is a shift by the lattice step $\Delta x$; any other shift on a given lattice is a power $\widehat{T}_{A} = (\widehat{T}_{\Delta x})^{A/\Delta x}$, where $A/\Delta x \in \mathbb{Z}_{N}$:
\begin{equation}\label{T1}
    \widehat{T}_{A}\psi(x) = \psi(x + A), \, \widehat{T}_A|x\rangle = |x - A\rangle, \, 
    \langle x^{\prime}|\widehat{T}_{A}|x^{\prime \prime}\rangle = \delta_{x^{\prime},x^{\prime \prime }- A} = \delta_{x^{\prime} + A,x^{\prime \prime}}.
\end{equation}
Moreover,
\begin{equation}\label{T2}
    \hat T_{\Delta x}\psi(x)=\hat T_{\Delta x}
\left(\begin{array}{c}
\psi((N-1)\Delta x)\\
\psi((N-2)\Delta x)\\
\vdots\\
\psi(2\Delta x)\\
\psi(\Delta x)\\
\psi(0)
\end{array}\right)=
\left(\begin{array}{c}
\psi(0)\\
\psi((N-1)\Delta x)\\
\psi((N-2)\Delta x)\\
\vdots\\
\psi(2\Delta x)\\
\psi(\Delta x)
\end{array}\right)=\psi(x+\Delta x).
\end{equation}
The sum $x + \Delta x$ is taken in the sense $x \in \Delta x \cdot \mathbb{Z}_{N}$, i.e., this is a cyclic shift of the function on the lattice down one position.
 
The eigenvalues of the minimum shift operator are $N$th roots of unity and are related to the eigenvalues of the momentum operator (which has not yet been introduced explicitly):
\begin{equation}
    \lambda^{N} = 1, \qquad \lambda_{p} = \E^{2\pi \I \Delta x p} =  \E^{2\pi \I \Delta x \Delta p p/\Delta p} = (\lambda_{\Delta p})^{p/\Delta p},
\end{equation}
where we take $\Delta x \Delta p = 1/N$ and $p/\Delta p \in \mathbb{Z}_{N}$ into account. The corresponding eigenvalues are obtained from the relation $\psi(x) = \widehat{T}_x\psi(0)$. The normalized eigenvectors have the forms
\bea
    \psi_{\lambda_{p}}(x^{\prime}) &=& \langle x^{\prime}|\psi_{\lambda_{p}}\rangle = \frac{\lambda_{p}^{x^{\prime}/\Delta x}}{\sqrt{N}} = \frac{\E^{2\pi \I x^{\prime} p}}{\sqrt{N}},\\ 
    \nonumber
    \langle \psi_{\lambda_{p}}|x^{\prime \prime}\rangle &=& \langle x^{\prime \prime}|\psi_{\lambda_{p}}\rangle^{*} = \frac{\lambda_{p}^{-x^{\prime \prime}/\Delta x}}{\sqrt{N}} = \frac{\E^{-2\pi \I x^{\prime \prime}}}{\sqrt{N}}.
\eea
We can write the projector on the (one-dimensional) eigensubspace of the operator $\widehat{T}_{\Delta x}$ as
\bea
    \widehat{P}_{\lambda_{p}} &=& |\psi_{\lambda_{p}}\rangle\langle\psi_{\lambda_{p}}|,\\ \nonumber
    \langle x^{\prime}|\widehat{P}_{\lambda_{p}}|x^{\prime \prime} \rangle &=& \frac{\lambda_{p}^{(x^{\prime} - x^{\prime \prime})/\Delta x}}{N} = \frac{\lambda_{p}^{d/\Delta x}}{N} = \frac{\E^{2\pi \I p d}}{N}, \quad d = x^{\prime} - x^{\prime \prime}.
\eea
In this notation $x^{\prime}$ labels rows of matrix, and $x^{\prime\prime}$ labels columns.
 
The eigenstates of the minimum shift operator are also eigenstates of the momentum operator and can be written differently:
\begin{equation}
    |\psi_{\lambda_{p}}\rangle = |\psi_{p}\rangle = |p\rangle, \quad \langle x | p \rangle = \frac{\E^{2\pi \I x p}}{\sqrt{N}}.
\end{equation}

\subsection{Group of shifts} 

We make a trivial remark that might nevertheless be of some interest for an arbitrary positional number system. 
We constructed the momentum operator such that it generates a symmetry group with respect to the shifts of the coordinate lattice by an integer number of nodes, i.e., a group isomorphic to the group (with respect to addition) of the residues modulo division by $N:\, \Delta x \cdot \mathbb{Z}_{N} \approx \mathbb{Z}_{N}$. But we can consider unitary operators of the form $\widehat{T}_{A} = e^{2\pi i A \hat{p}}, \, A \in \mathbb{R}.$ Such operators correspond to the cyclic shifts by an arbitrary value (not necessarily a multiple of $\Delta x$). The corresponding group is isomorphic to the group $\mathbb{R} / (\Xi \cdot \mathbb{Z}) \approx SO(1) \approx U(1)$ of a circle’s rotations by an arbitrary angle. Addition is again understood in terms of modulo $\Xi$, ($A = A + \Xi$). In the case $\Xi = \infty$, the group of symmetries coincides with the group $\mathbb{R}$ of real numbers with respect to addition.
 
We see that if the Hamiltonian on the lattice is expressed in terms of the operator $\hat{p}$, then the presence of the lattice does not violate translation invariance under arbitrary translations (not necessarily by an integer number of the lattice sites), but the operator $\hat{p}$ (as we see below) turns out to be nonlocal, i.e., matrix elements $\langle x^{\prime} | \hat{p} | x^{\prime\prime} \rangle$ can be nonzero for arbitrary large values $x^{\prime} - x^{\prime \prime}$ (in the lattice).
 
We can specify a state $|x_{0}\rangle = \widehat{T}_{x_{0}}|0\rangle$ with an arbitrary value $x_{0} \notin \Delta x \cdot \mathbb{Z}_{N}$, but such a state is not a state with a certain value of the coordinate because it decomposes into several basic states $\{ |x\rangle \}_{x \in \Delta x \cdot \mathbb{Z}_{N}}$.

\section{Operators of digits and their decomposition by shifts}
\subsection{Operators of digits on the lattice}
We defined the momentum operator such that the Fourier harmonic of the momentum is given by the operator $\widehat{T}_{A} = \E^{2\pi \I A \hat{p}}$ of the coordinate shift. Therefore, if we take Fourier transform for the momentum digits
\begin{equation}
    \mathbf{t}_{n_{+}n_{-}}(r,p) = \sum\limits_{A \in \Delta x \cdot \mathbb{Z}_{N}}\tilde{\mathbf{t}}_{n_{+}n_{-}}(r, A)\E^{2\pi \I A p},
\end{equation}
then we obtain the decomposition of the momentum digit by coordinate shifts%
\begin{equation}
   \mathbf{t}_{n_{+}n_{-}}(r, \hat{p}) = \sum\limits_{A \in \Delta x \cdot \mathbb{Z}_{N}} \tilde{\mathbf{t}}_{n_{+}n_{-}}(r, A) \widehat{T}_{A}.
\end{equation}
%Здесь начинается импровизация, этот кусок надо будет тщательно проверить.
 
In this article, we are going to discuss two possible representations of ternary digits -- with values $\{0,1,-1\}$ (we will call this representation "symmetric system") and $\{0,1,2\}$ (a "non-symmetric system"). 
%We are going to discuss them separately and than compare the results.(Это не соответствует написанному ниже)\\
\subsubsection{Operators of digits on the lattice for the symmetric system}
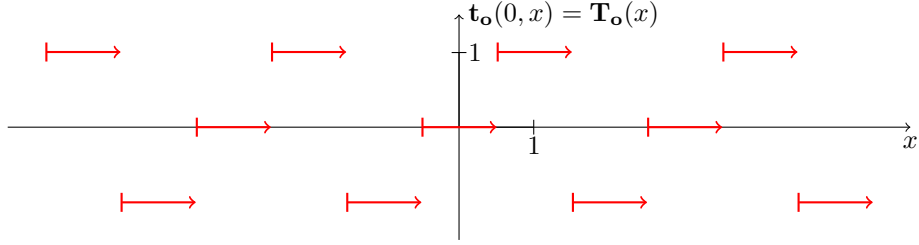
\begin{figure}[h]
    \centering
\begin{tikzpicture}
\draw [->] (-6,0) -- (6,0) node [below] {$x$};
\draw [-|] (0,0) -- (1,0) node [below] {1};
\draw [-|] (0,0) -- (0,1) node [right] {1};
\draw [->] (0,-1.5) -- (0,1.5) node [right] {$\mathbf{t_o}(0,x)=\mathbf{T_o}(x)$};

\draw[|->,thick,red] (-0.5,0)--(0.5,0);
\draw[|->,thick,red] (0.5,1)--(1.5,1);
\draw[<-|,thick,red] (-0.5,-1)--(-1.5,-1);

\draw[|->,thick,red] (1.5,-1)--(2.5,-1);
\draw[|->,thick,red] (2.5,0)--(3.5,0);
\draw[|->,thick,red] (3.5,1)--(4.5,1);

\draw[|->,thick,red] (4.5,-1)--(5.5,-1);

\draw[<-|,thick,red] (-1.5,1)--(-2.5,1);
\draw[<-|,thick,red] (-2.5,0)--(-3.5,0);
\draw[<-|,thick,red] (-3.5,-1)--(-4.5,-1);

\draw[<-|,thick,red] (-4.5,1)--(-5.5,1);
\end{tikzpicture}\\
    \caption{Plot of the value of the \textbf{t}ernary digit before the ternary point (multiplier with $3^{0}$) for a ``symmetric system'' is periodic and \textbf{o}dd (except for the discontinuity points).}
\end{figure}
After simple computations (see the appendix), we obtain the expansion of the operator $\hat{p}_{r}=\mathbf{t_o}(r,\hat p)$ of the momentum digit of the particle on the lattice over the shift operators $\widehat{T}_{A}$:
\begin{equation}
    \hat{p}_r =  \Delta p\, 3^{-r}\sum\limits_{D \in \mathbb{Z}_{3^r/\Delta p}} \sum\limits_{\sigma = 1}^{2}  \frac{(-1)^{D + \sigma}}{2 \I \sin\left(\pi\,\Delta p \,A\right)}\widehat{T}_{-A},\quad 
    A=3^{-r}(D + \sigma/3).
\end{equation}
Similarly, the operator $\hat x_s=\mathbf{t_o}(s,\hat x)$ of the coordinate digit can be expanded in the momentum shifts $\widehat{S}_{B} = e^{-2\pi i \hat{x} B}$:
\begin{equation}
     \hat{x}_s =  \Delta x\,3^{-s}\sum\limits_{D \in \mathbb{Z}_{3^s/\Delta x}} \sum\limits_{\sigma = 1}^{2}  \frac{(-1)^{D + \sigma}}{2 \I \sin\left(\pi\,\Delta x\,B\right)}\widehat{S}_{B},\quad
     B=3^{-s} (D + \sigma/3).
\end{equation}
\subsubsection{Operators of digits on the lattice for the non-symmetric system}
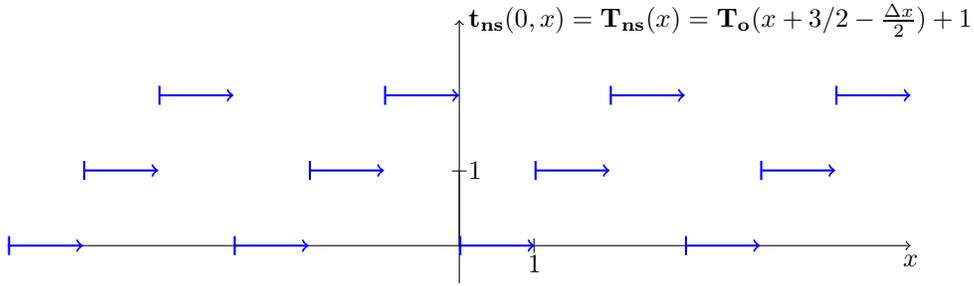
\begin{figure}[h!]
    \centering
\begin{tikzpicture}
\draw [->] (-6,0) -- (6,0) node [below] {$x$};
\draw [-|] (0,0) -- (1,0) node [below] {1};
\draw [-|] (0,0) -- (0,1) node [right] {1};
\draw [->] (0,-0.5) -- (0,3) node [right] {$\mathbf{t_{ns}}(0,x)=\mathbf{T_{ns}}(x)=\mathbf{T_o}(x+3/2 - \frac{\Delta x}{2})+1$};

\draw[|->,thick,blue] (0,0)--(1,0);
\draw[|->,thick,blue] (1,1)--(2,1);
\draw[|->,thick,blue] (2,2)--(3,2);
\draw[|->,thick,blue] (3,0)--(4,0);
\draw[|->,thick,blue] (4,1)--(5,1);
\draw[|->,thick,blue] (5,2)--(6,2);

\draw[|->,thick,blue] (-6,0)--(-5,0);
\draw[|->,thick,blue] (-5,1)--(-4,1);
\draw[|->,thick,blue] (-4,2)--(-3,2);
\draw[|->,thick,blue] (-3,0)--(-2,0);
\draw[|->,thick,blue] (-2,1)--(-1,1);
\draw[|->,thick,blue] (-1,2)--(0,2);
\end{tikzpicture}
    \caption{Plot of the value of the \textbf{t}ernary digit before the ternary point (multiplier with $3^{0}$) for a ``\textbf{n}on-\textbf{s}ymmetric system'' is periodic.}
\end{figure}
The same computations for the non-symmetric system give us the following expressions for the expansion of the operator $\hat{p}_{r}=\mathbf{t_{ns}}(r,\hat p)$:
\begin{equation}
     \hat{p}_{r} = \hat{1} -
     \Delta p\, 3^{-r}\sum\limits_{D \in \mathbb{Z}_{3^{r}/\Delta p}} \sum\limits_{\sigma = 1}^{2} 
     \frac{\widehat{T}_{-A}}{1 - \exp\left(2\pi \I\, \Delta p\, A\right)},\quad
     A=3^{-r}(D + \sigma /3)
\end{equation}
and the operator of the coordinate digit $\hat x_s=\mathbf{t_{ns}}(s,\hat x)$:
\begin{equation}
\hat{x}_{s} = \hat{1} -\Delta x\,3^{-s}\sum\limits_{D \in \mathbb{Z}_{3^{s}/\Delta x}} \sum\limits_{\sigma = 1}^{2} 
\frac{\widehat{S}_{B}}{1 - \exp\left(2\pi \I\, \Delta x\,B\right)},
\quad
B=3^{-s} (D + \sigma /3).
\end{equation}
\subsubsection{Operators of digits on the lattice for binary systems}
Here, we compare the results, obtained for the ternary case with those, which are discussed  for the binary case, considered in \cite{bin-ivanov}, which we call ``binary non-symmetric system.''

The coordinate lattice is $\Delta x\cdot\mathbb{Z}_{2^n}$ ($\Delta x=2^{-n_-}$).

The momentum lattice is $\Delta p\cdot\mathbb{Z}_{2^n}$ ($\Delta p=2^{-n_+}$).

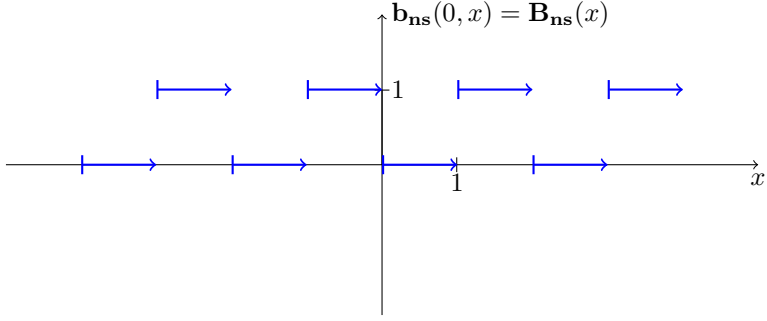
\begin{figure}[h]
\centering
\begin{tikzpicture}
\draw [->] (-5,0) -- (5,0) node [below] {$x$};
\draw [-|] (0,0) -- (1,0) node [below] {1};
\draw [-|] (0,0) -- (0,1) node [right] {1};
\draw [->] (0,-2) -- (0,2) node [right] {$\mathbf{b_{ns}}(0,x)=\mathbf{B_{ns}}(x)$};

\draw[|->,thick,blue] (0,0)--(1,0);
\draw[|->,thick,blue] (1,1)--(2,1);
\draw[|->,thick,blue] (2,0)--(3,0);
\draw[|->,thick,blue] (3,1)--(4,1);

\draw[|->,thick,blue] (-4,0)--(-3,0);
\draw[|->,thick,blue] (-3,1)--(-2,1);
\draw[|->,thick,blue] (-2,0)--(-1,0);
\draw[|->,thick,blue] (-1,1)--(0,1);
\end{tikzpicture}
\caption{Plot of the value of the \textbf{b}inary digit before the binary point (multiplier with $2^{0}$) for an ``\textbf{n}on-\textbf{s}ymmetric system.''\label{fig-bin-ns}}
\end{figure}

The \textbf{b}inary digit for the \textbf{n}on-\textbf{s}ymmetric system we denote as $\hat p_r=\mathbf{b_{ns}}(r,\hat p)$
%For the even system in \cite{bin-ivanov} was obtained the following result:
\bea
    \mathbf{b_{ns}}(r,\hat{p}) &=& \frac{\hat{1}}{2} - \Delta p\,2^{-r}\sum\limits_{D \in \mathbb{Z}_{2^r/\Delta p}} \frac{\widehat{T}_{-A}}{1 - \exp(2\pi \I \Delta p\, A )}=\\ \nonumber
    &=&\frac{\hat{1}}{2}+ \Delta p\,2^{-r}\sum\limits_{D \in \mathbb{Z}_{2^{r}/\Delta p}} 
    \frac{ \exp\left(-\pi \I \,\Delta p\,A\right)}{2\I \sin\left(\pi \I\, \Delta p\,A\right)}
    \widehat{T}_{-A},\\
    &&A=2^{-r}(D + 1/2),\quad \Delta p=2^{-n_+}.\nonumber
\eea

By comparing with the ternary case, we found that a binary ``symmetric'' system with the digits $x_s \in \{\pm 1/2\}$ is also possible.

The coordinate lattice and the momentum lattice are shifted by one half of the lattice constants $\Delta x=2^{-n_-}$ and $\Delta p=2^{-n_+}$, respectively. Zero is no longer a point of the lattices!

The coordinate lattice is $\Delta x\cdot(\frac12+\mathbb{Z}_{2^n})
=\{\frac{\Delta x}{2},\frac{3\Delta x}{2},\frac{5\Delta x}{2},\dots,(2^n-\frac12)\Delta x\}$.

If we consider the shift operator of the form \eqref{T1}, \eqref{T2}, then we have 
$p\in\Delta p\cdot\mathbb{Z}_{2^n})$, which cannot be represented by the  symmetric binary system.

To use the symmetric binary system for the momentum, we can impose the antiperiodic boundary
condition $\psi(x+2^{n_+})=-\psi(x)$, the corresponding minimal shift operator has the following form
\begin{equation}
	\hat T^-_{\Delta x}\psi(x)=\hat T^-_{\Delta x}
	\left(\begin{array}{c}
		\psi((N-1)\Delta x)\\
		\psi((N-2)\Delta x)\\
		\vdots\\
		\psi(2\Delta x)\\
		\psi(\Delta x)\\
		\psi(0)
	\end{array}\right)=
	\left(\begin{array}{c}
		-\psi(0)\\
		\psi((N-1)\Delta x)\\
		\psi((N-2)\Delta x)\\
		\vdots\\
		\psi(2\Delta x)\\
		\psi(\Delta x)
	\end{array}\right),
\end{equation}
where $N=2^n$.

The momentum lattice is $\Delta p\cdot(\frac12+\mathbb{Z}_{2^n})
=\{\frac{\Delta p}{2},\frac{3\Delta p}{2},\frac{5\Delta p}{2},\dots,(2^n-\frac12)\Delta p\}$.

\bea
\mathbf{b_o}(r,\hat{p}) 
%&=& -\Delta p \, 2^{-r} \sum\limits_{D \in \mathbb{Z}_{2^{r}/\Delta p}} \frac{\widehat{T}_{-A}}{1 - \exp(2\pi \I \Delta p A)}\exp(2\pi \I \Delta p A) =\\ \nonumber
%&=& \Delta p \, 2^{-r} \sum\limits_{D \in \mathbb{Z}_{2^{r}/\Delta p}} \frac{\widehat{T}_{-A}}{1 - \exp(-2\pi \I \Delta p A)} \qquad  =\\ \nonumber
%&=&  \Delta p \, 2^{-r}  \sum\limits_{D \in \mathbb{Z}_{2^{r}/\Delta p}} \frac{\exp(\pi \I \Delta p A)}{2 \I \sin (\pi \I \Delta p A)}\widehat{T}_{-A}, \\
&=&\mathbf{b_{ns}}\left(r,\hat p-\frac{\Delta p}{2}\right)-\frac{\hat 1}2=
%&=&  
\Delta p \, 2^{-r}  \sum\limits_{D \in \mathbb{Z}_{2^{r}/\Delta p}} \frac{\widehat{T}_{-A}}{2 \I \sin (\pi \I \Delta p A)}, \\
 &&A=2^{-r}(D + 1/2),\quad \Delta p=2^{-n_+}.\nonumber
\eea

\begin{figure}[h]
\centering
\begin{tikzpicture}
\draw [->] (-5,0) -- (5,0) node [below] {$x$};
\draw [-|] (0,0) -- (1,0) node [below] {1};
\draw [-|] (0,0) -- (0,1) node [right] {1};
\draw [->] (0,-2) -- (0,2) node [right] {$\mathbf{b_o}(0,x)=\mathbf{b_{ns}}(0,x-\tfrac{\Delta x}{2})-\frac12$};

\draw[|->,thick,red] (0,-0.5)--(1,-0.5);
\draw[|->,thick,red] (1,0.5)--(2,0.5);
\draw[|->,thick,red] (2,-0.5)--(3,-0.5);
\draw[|->,thick,red] (3,0.5)--(4,0.5);

\draw[|->,thick,red] (-4,-0.5)--(-3,-0.5);
\draw[|->,thick,red] (-3,0.5)--(-2,0.5);
\draw[|->,thick,red] (-2,-0.5)--(-1,-0.5);
\draw[|->,thick,red] (-1,0.5)--(0,0.5);
\end{tikzpicture}
\caption{Plot of the value of the \textbf{b}inary digit before the binary point (multiplier with $2^{0}$) for the ``symmetric binary system''  is periodic and \textbf{o}dd (except for the discontinuity points).\label{fig-bin-o}}
\end{figure}
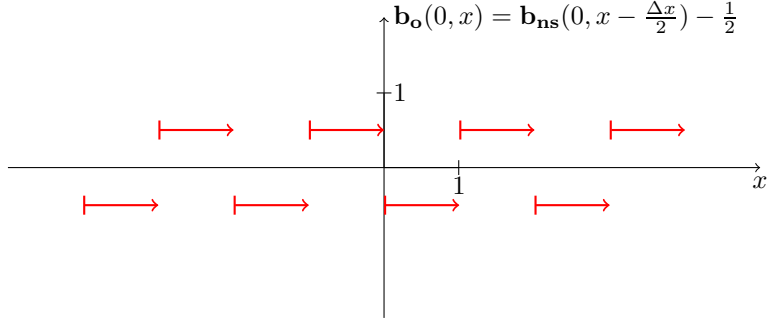

\subsection{Ternary expansion on the line}
We considered the case where the coordinate and momentum are given on a finite lattice. 
We consider the transition to a continuous limit. 
As we see below, this transition is associated with a nontrivial generalization of the sum of the power series.

In the case of the non-symmetric binary or the ternary system ($q=2$, $c=\mathbf{b_{ns}}$ or $q=3$, $c=\mathbf{t_{ns}}$)
the power series
$$
  \sum_{s=-\infty}^{+\infty}c(s,x)\,q^s,
$$
converge only for positive $x$ and diverge for any negative $s$.
$$
  \forall x>0,~s>\log_q x:\qquad  c(s,x)=0,
$$
$$
  \forall x<0,~s>\log_q (-x):\quad  c(s,x)=q-1.
$$

For the symmetric binary system, the system the power series diverge for any real number
$$
  \forall x>0,~s>\log_2 x:\qquad  \mathbf{b_{o}}(s,x)=-1/2,
$$
$$
  \forall x<0,~s>\log_2 (-x):\quad  \mathbf{b_{o}}=+1/2.
$$

We need some modification of power series, which is finite for any real $x$.
We will call the corresponding modification of the summation rule a \textit{renormalization}.

%\subsection{Renormalisation of ternary sums in non-symmetric system} 

\subsubsection{Renormalisation on the real line}

To assign a finite value to the ternary expansion of the negative number $x<0$, we must renormalize it, which reduces to assigning a finite value to the sum of the divergent geometric progression using the formula $\sum\limits_{s =0}^{+\infty} q^{s} = 1/(1-q)$ beyond its applicability limits: for $q = 3$. The renormalized sum is denoted by the sum symbol with a prime:
\begin{equation}
    \sum_{s = 0}^{~~\infty ~\prime} 3^{s} = \frac{1}{1 - 3} = -\frac{1}{2}.
\end{equation}
The same rule can be reduced to the form
\begin{equation}
    \sum_{s\in\mathbb{Z}}^{~~~~\prime}3^{s} = 0.
\end{equation}
We could consider the same series in the 3-adic sense \cite{p-adic}. However, we lose the convergence of fractions with an infinite number of nonzero digits after the ternary point and obtain the convergence of fractions with infinite digits before the ternary point. Hence, 3-adic convergence proves inconvenient for real coordinates because it must be applied selectively (only for the integer part of negative numbers).
Nevertheless, we can consider 3-adic observables, also decomposing them in ternary digits.

To apply the ternary expansion to the operators $\hat{x}$ and $\hat{p}$, the general formula for negative and positive numbers is useful. It is easily obtained using the formal computations
\begin{equation}
    x = \frac{3x - x}{2} = \frac{1}{2}\sum_{s\in\mathbb{Z}}^{~~~~\prime}(x_{s-1} - x_{s})3^{s}.
\end{equation}
It is easy to see that such a series converges to $x$ regardless of the sign because in both cases for the higher powers $x_{s-1} - x_{s} = 0$. The same rule ensures that the series is terminated in negative powers if all digits following the ternary point are the same after a certain position. Hence, the renormalized sum (with a prime) has the form
\begin{equation}
    x = \sum_{s\in\mathbb{Z}}^{~~~~\prime}x_s 3^{s} = \frac{1}{2}\sum_{s\in\mathbb{Z}}(x_{s-1} - x_{s}) \cdot 3^{s}.
\end{equation}
The binary version of this representation is known in computer science as the signed-digit representation \cite{Avizienis}, \cite{ChowRobertson}. It was introduced to increase the computational speed by reducing the number of carries of digits.
For the operator $\hat{x}$ and its digits, we similarly have
\begin{equation}
    \hat{x} = \sum_{s\in\mathbb{Z}}^{~~~~\prime}\hat{x}_s3^s 
    = \frac{1}{2} \sum_{s\in\mathbb{Z}}(\hat{x}_{s - 1} - \hat{x}_{s})\cdot 3^{s}.
\end{equation}
The operator of digit $\hat x_{s} = \mathbf{t_{ns}}(s, \hat{x})$ is a self-adjoint operator with the eigenvalues $0,1,2$.

In the binary case \cite{bin-ivanov}, we have a similar renormalized sum for numbers and operators
\begin{equation}\label{bin-ren}
\hat{x}=\sum_{s\in\mathbb{Z}}^{~~~~\prime}\hat{x}_s2^s =\sum_{s\in\mathbb{Z}}(\hat{x}_{s - 1} - \hat{x}_{s})\cdot 2^{s},
\end{equation}
where the operator of digit $\hat x_{s} = \mathbf{b_{ns}}(s, \hat{x})$ is a self-adjoint operator with the eigenvalues $0,1$.

%\subsubsection{Binary symmetric system} %Вроде исправил
Here, we need to mention that the binary symmetric system on the real line can be obtained from a non-symmetric one with substitution $0 \to -1/2$, $1 \to +1/2$. Thus for positive numbers, all digits after a particular one are $-1/2$, for negative -- $+1/2$. Hence, the sum diverges for both positive and negative numbers on the line, and so renormalization is necessary in both cases:
\begin{equation}
x = 2x - x = \sum_{s\in\mathbb{Z}}^{~~~~\prime}x_s2^s = \sum_{s\in\mathbb{Z}}(x_{s-1} - x_{s}) \cdot 2^{s}.
\end{equation}
The renormalized sum on the real line is identical for symmetric and non-symmetric expansion because in this case
(see Fig. \ref{fig-bin-ns}, \ref{fig-bin-o})
$\mathbf{b_{o}}=\mathbf{b_{ns}}-\frac12$.

\subsubsection{Renormalization on the lattice} 
 
We can also generalize the renormalization method to the case where the coordinate is defined on the lattice.

There are no infinite nodes on the lattice $\Delta x\cdot\mathbb{Z}_N$.
The renormalization on the lattice is a change of representation of $\mathbb{Z}_N$ from
$\{0,1,2,\dots, N-1\}$ to $\{-k,-k+1,\dots,-2,-1,0,1,2,\dots,N-k-1\}$.

In the binary case \cite{bin-ivanov}, we can just reduce the range of summation in \eqref{bin-ren}:
\begin{equation}\label{method1}
	x^{\prime} = \sum\limits_{s = -n_{-}}^{n_{+} - 1~\prime}x_s 2^s = 
	\sum\limits_{s = -n_{-}}^{n_{+} - 1} (x_{s-1} - x_s)2^{s}, \quad x_{-n_{-} - 1} = 0.
\end{equation}
The same renormalization could be represented by redefinition of the last digit
$x'_{n_+-1}=-x_{n_+-1}$:
\begin{equation}\label{method2}
\sum\limits_{s = -n_{-}}^{n_{+} - 1~\prime}x_s 2^s
= \sum\limits_{s = -n_{-}}^{n_{+} - 2}x_s 2^s +x'_{n_+-1}2^{n_+-1}.
\end{equation}
The binary renormalization on the lattice is linear with respect to the binary digits $x_s$.

In the ternary case, the two methods \eqref{method1}, \eqref{method2} are no longer equivalent.

The second method \eqref{method2} works. It changes the lattice 
$\{0,\Delta x,2\Delta x,3\Delta x,\dots, (3^n-1)\Delta x\}$
to the lattice
\\ $\{-3^{n-1}\Delta x,\dots,-\Delta x,0,\Delta x,2\Delta x,\dots,(2 \cdot 3^{n-1}-1)\Delta x\}$ by
substracting $3^{n_+}$ from the last $3^{n-1}$ nodes.
\bea\label{method2a-3}
  \mathbf{t'_{ns}}(n_+-1,x)&=&\left\{\begin{array}{cc}
   0,&\mathbf{t_{ns}}(n_+-1,x)=0,\\
   1,&\mathbf{t_{ns}}(n_+-1,x)=1,\\
   -1,&\mathbf{t_{ns}}(n_+-1,x)=2
  \end{array}\right.=\\ \nonumber
  &=&\mathbf{t_{ns}}(n_+-1,x)-\frac32(\mathbf{t_{ns}}(n_+-1,x)-1)\mathbf{t_{ns}}(n_+-1,x).
\eea
\begin{equation}\label{method2b-3}
\sum\limits_{s = -n_{-}}^{n_{+} - 1~\prime}\mathbf{t_{ns}}(s,x) 3^s 
= \sum\limits_{s = -n_{-}}^{n_{+} - 2}\mathbf{t_{ns}}(s,x)\, 3^s +\mathbf{t'_{ns}}(n_+-1,x)\, 3^{n_+-1}.
\end{equation}
The ternary renormalization on the lattice \eqref{method2a-3}, \eqref{method2b-3} is not linear with respect to the ternary digits $x_s$.

The first method \eqref{method1} does not work.
The correspondig sum has the form
\begin{equation}
    x^{\prime\prime} = \sum\limits_{s = -n_{-}}^{n_{+} - 1~\prime\prime}\mathbf{t_{ns}}(s,x) 3^s = 
    \frac{1}{2}\sum\limits_{s = -n_{-}}^{n_{+} - 1} (\mathbf{t_{ns}}(s - 1, x) - \mathbf{t_{ns}}(s,x))3^{s}, \quad \mathbf{t_{ns}}(-n_{-} - 1, x) = 0.
\end{equation}
The maximal ternary number is renormalized in the appropriate way
\begin{equation}
    \sum\limits_{s = -n_{-}}^{n_{+} - 1 ~\prime\prime}2\cdot 3^s = \frac{1}{2} (-2\cdot 3^{-n_{-}}) =-\Delta x.
\end{equation}
One-half of the previous number is out of the lattice
\begin{equation}\label{polovina}
\sum\limits_{s = -n_{-}}^{n_{+} - 1 ~\prime\prime} 3^s =-\frac{\Delta x}{2}.
\end{equation}

Any renormalization on the lattice assume the mapping 
$$
0\mapsto0,\qquad 
(N-1)\Delta x\mapsto-\Delta x.
$$
If renormalization is linear with respect to the ternary digits $x_s$, then $\frac{N-1}{2}\Delta x\mapsto-\frac{\Delta x}{2}$, like \eqref{polovina}, but $-\frac{\Delta x}{2}$ is out of the lattice.

Any ternary renormalization on the lattice is not linear with respect to the ternary digits $x_s$.

\subsection{Integral ternary representation}
\subsubsection{Representation for positive numbers}
The function $\mathbf{t}(s,x)$ is expressed in terms of a function of one variable 
$\mathbf{T}(x) = \mathbf{t}(0,x)$ by the formula 
\begin{equation}
\mathbf{t}(s,x) = \mathbf{t}(0, 3^{-s}x) = \mathbf{T}(3^{-s}x).
\end{equation}
On the real line, we can consider noninteger values of $s$, which can be useful, for example, when changing the scale of a unit interval. Then $x \to 3^{a}x$ and $\mathbf{t}(s,x) \to \mathbf{t}(s-a,x)$. For positive numbers $\mathbf{t}(s,x) = \mathbf{t}(s - \log_{3} x, 1)$, $x > 0$. For negative numbers $\mathbf{t}(s,x) = Z - \mathbf{t}(s, |x|) = Z - \mathbf{t}(s - \log_{3}|x|,1)$, $x < 0$, $Z=0$ for $\mathbf{t_o}$ and $Z=2$ for $\mathbf{t_{ns}}$.
 
By analogy with the ternary series, we consider the formal ternary integral
\begin{equation}
    \int\limits_{-\infty}^{+\infty}\mathbf{t}(s,x)3^{s} ds.
\end{equation}

For any $x>0$, the highest digits are zero, $\mathbf{t}(s,x) = 0$ for all $s > \log_{3}x$. The ternary integral
\begin{equation}
\int\limits_{-\infty}^{+\infty}\mathbf{t}(s,x)3^{s}ds = \int\limits_{-\infty}^{\log_{3}x}\mathbf{t}(s,x)3^s ds
\end{equation}
converges because the integrand tends to zero as $s \to +\infty$ and is majorized by a convergent integral of the exponential function as $s \to -\infty$.

For positive numbers, it is easy to prove by a direct calculation that the ternary integral, like the ternary sum, gives $x$:
\begin{equation}
\int\limits_{-\infty}^{+\infty}\mathbf{t}(s,x)3^{s} ds
=\int\limits_{-\infty}^{+\infty}\mathbf{T}(3^{-s}x) \frac{d3^{s}}{\ln 3}
=\int\limits_{0}^{+\infty}\mathbf{T}(3^{-s}x) \frac{d3^{s}}{\ln 3}
=x\cdot\int\limits_{0}^{+\infty}\mathbf{T}(S^{-1}) \frac{dS}{\ln 3},
\end{equation}
where $S=3^{s}/x$. Now we need to check that
$$
\frac{1}{\ln 3}\int\limits_{0}^{+\infty}\mathbf{T}(S^{-1})\,dS=1.
$$

There is a more obvious proof.
One can rescale $x$ 
\bean
    x3^{-\Delta s} = \sum_{s\in\mathbb{Z}}\mathbf{t}(s,x3^{-\Delta s})3^s
    = \sum_{s\in\mathbb{Z}}\mathbf{t}(s+\Delta s,x)3^s~\Rightarrow~\\
    ~\Rightarrow~
    x=\sum_{s\in\mathbb{Z}}\mathbf{t}(s+\Delta s,x)3^{s+\Delta s}.
% = \int\limits_{-\infty}^{+\infty}\mathbf{t}(s,x)3^{s} ds.
\eean
%We can also introduce the integral representation in a more natural way.
 So we can consider $x$ to be expanded as a raw of digits in non-integer nodes with a unit step:
\begin{equation}
    x = \sum\limits_{s \in \mathbb{Z}}x_{s + \Delta s}3^{s + \Delta s}. 
%    = \int\limits_{0}^{1}\underbrace{x}_{\text{const.}}d(\Delta s)
\end{equation}
Hence, we obtain the following expression:
\begin{equation}
    x=\int\limits_{0}^{1}\underbrace{x}_{\text{const.}}d(\Delta s) 
    = \int\limits_{0}^{1}\left(\sum\limits_{s\in \mathbb{Z}}x_{s + \Delta s}3^{s + \Delta s}\right) d(\Delta s)
\end{equation}
If the sum converges uniformly, we can swap the integral with the sum and obtain the integral representation:
\begin{equation}
     x = \sum\limits_{s \in \mathbb{Z}} \int\limits_{0}^{1}x_{s + \Delta s}3^{s + \Delta s} d(\Delta s) 
     %= \bigg{|} s + \Delta s = \mathscr{S} \bigg{|} 
     = \sum\limits_{s \in \mathbb{Z}}\int\limits_{s}^{s + 1}x_{\mathscr{S}}3^{\mathscr{S}}d\mathscr{S} = \int\limits_{-\infty}^{\infty}x_{\mathscr{S}}3^{\mathscr{S}}d\mathscr{S},
\end{equation}
where $\mathscr{S}=s+\Delta s$.

\subsubsection{Renormalization for non-symmetric system}
For any $x<0$ in the non-symmetric system, the highest digits are equal to $2$, $\mathbf{t_{ns}}(s,x) = 2$ for all $s \geqslant \log_{3}|x|$. The ternary integral
\begin{equation}
    x = \int\limits_{-\infty}^{+\infty \,\,\prime}\underbrace{\mathbf{t_{ns}}(s,x)}_{2-\mathbf{t_{ns}}(s,|x|)}3^s\,ds = \int\limits_{-\infty}^{+\infty \,\, \prime}3^{s}\,ds
    -\underbrace{\int\limits_{-\infty}^{+\infty}\mathbf{t_{ns}}(s, |x|)3^{s}\,ds}_{|x|} 
    =\int\limits_{-\infty}^{+\infty \,\, \prime}3^sds+x
\end{equation}
diverges because a divergent summand is added to the convergent integral.
 
To assign a finite value to the ternary integral representation of a negative number $x < 0$, we should renormalize it. Similarly to the ternary sum, the renormalization corresponds to applying the formula $\int\limits_{0}^{\infty}q^sds = -1/\log q$, which holds for $0<q<1$, to the case $q > 1$:
\begin{equation}
    \sum_{s\in\mathbb{Z}}3^s = \int\limits_{s=-\infty}^{+\infty \,\, \prime} 3^s ds = 0 \quad \Longleftrightarrow  \quad
    \sum_{s\in\mathbb{Z}}^{~~~~~\prime}3^s = \log_3 \int\limits_{0}^{+\infty \,\, \prime} 3^s ds = -1.
\end{equation}
As for the ternary sum, we can obtain the general formula that converges regardless of the sign of $x$:
\begin{equation}
    x=\frac{3x-x}{2} 
    =\frac{1}{2}\int\limits_{-\infty}^{\infty}\left(\mathbf{t_{ns}}(s-1,x)-\mathbf{t_{ns}}(s,x)\right)3^s ds.
\end{equation}
The other way of renormalization is to integrate by parts and omit the divergent summand:
\begin{equation}
    \int\limits_{-\infty}^{\infty \, \prime \prime} \mathbf{t_{ns}}(s,x) 3^s ds = \underbrace{\frac{1}{\ln 3}\mathbf{t_{ns}}(s,x)3^s\bigg|^{+\infty\prime}_{-\infty}}_\text{0 by definition} - \frac{1}{\ln 3} \int\limits_{-\infty}^{\infty } (\mathbf{t_{ns}}(s,x))^{\prime} 3^s ds = x.
\end{equation}
The derivative $(\mathbf{t_{ns}}(s,x))^{\prime}=\frac{d~}{ds}\mathbf{t_{ns}}(s,x)$ is understood in the sense of generalized functions.

The integral representation is overdetermined because to specify a number, it suffices to specify its digits with integer numbers or digits with a unit step, which corresponds to the choice of $3^{\Delta s}$ as a scale segment:  for any $\Delta s \in \mathbb{R}$,
\begin{equation}
    x = \sum\limits_{s\in\mathbb{Z}}\mathbf{t_{ns}}(s + \Delta s, x) 3^{s + \Delta s} = 3^{\Delta s}\sum\limits_{s\in\mathbb{Z}}\mathbf{t_{ns}}(s, x\cdot3^{-\Delta s})3^s.
\end{equation}
Such a representation has its advantages despite being overdetermined. In particular, the integral representation is scale invariant: it does not distinguish scales of the form $3^{s}$, $s \in \mathbb{Z}$.

\subsection{Operators of digits on the line}
If we take the limit $n \to +\infty$ for $\Delta p \to 0$ in expansions for the operators of digits on the lattice, we can obtain the following expansions on the line:
\bea
    \mathbf{t_{o}}(r,\hat p)
    &=&- \sum\limits_{D \in \mathbb{Z}} \sum\limits_{\sigma = 1}^{2} 
    \frac{(-1)^{D+\sigma}\widehat{T}_{-3^{-r}(D + \sigma/3)}}{2\pi \I (D + \sigma/3)}, 
    %\E^{\pi \I (3D + \sigma)},
                \\
    \mathbf{t_{ns}}(r,\hat p)
    &=& \hat{1} + \sum\limits_{D \in \mathbb{Z}} \sum\limits_{\sigma = 1}^{2}\frac{\widehat{T}_{-3^{-r}(D + \sigma/3)}}{2\pi \I (D + \sigma/3)}.
\eea
This corresponds to an infinite number of digits before the ternary point for the coordinate and an infinite number of digits after the ternary point  for the momentum, i.e., to a lattice in the coordinate and a circle in the momentum (if also $\Delta x \to 0$, then both the coordinate and the momentum yield a line). Specifically, we note that we obtain a sum over all integers, including negative integers, in the limit. Before passing to the limit for the lattice of finite size closed to a circle, separating numbers into negative and positive did not make sense. The same expressions can be obtained after direct Fourier transform of respective functions on the line.
 
For binary systems, we obtain the following expression:
\begin{equation}
    \mathbf{b}(r,\hat p)=
    %\hat{p}_{r} = 
    \sum\limits_{D \in \mathbb{Z}}\frac{\widehat{T}_{-2^{-r}(D + 1/2)}}{2\pi \I (D + 1/2)}+
    \left\{\begin{array}{cr}
    	0,&\text{symmetric system}\\
	    \frac{\hat 1}{2},&\text{non-symmetric system}\\
    	\end{array}\right..
\end{equation}
\section{Commutation relations}
\subsection{Digit-digit commutator}
The operator of the momentum digits is expanded in terms of shifts operators:
\bea
 A&=&3^{-r}(D + \sigma /3)\\
  \mathbf{t_{o}}(r,\hat p)&=&
  %\hat{p}_{r_{sym}} = 
  - \Delta p \cdot 3^{-r} \,\sum\limits_{D \in \mathbb{Z}_{3^r/\Delta p}} \sum\limits_{\sigma = 1}^{2} \frac{(-1)^{D+\sigma}\widehat{T}_{-A}}{2\I \sin\left(\pi \Delta p A\right)},
           \\
  \mathbf{t_{ns}}(r,\hat p)&=&
  %\hat{p}_{r_{n.s.}} = 
  \widehat{1} - \Delta p \cdot 3^{-r} \,\sum\limits_{D \in \mathbb{Z}_{3^{r}/\Delta p}} \sum\limits_{\sigma = 1}^{2} \frac{\widehat{T}_{-A}}{1 - \exp\left(2\pi \I\, \Delta p\, A\right)}.
\eea
%Here and hereafter the operator with index "sym" corresponds to the symmetric system and with index "n.s." -- to a non-symmetric one.
It is easy to derive the commutation relation between the arbitrary function of the coordinate $f(\hat{x})$ and the shift operator $\widehat{T}_{-A}$:
%\begin{equation}
%    [f(\widehat{x}), \widehat{T}_{A}]\psi(x) = (f(\widehat{x}) - f(\widehat{x} + A))\widehat{T}_{A}\psi(x).
%\end{equation}
%Hence,
\begin{equation}
    [f(\widehat{x}), \widehat{T}_{-A}] = (f(\widehat{x}) - f(\widehat{x} - A))\widehat{T}_{-A}.
\end{equation}
Because the digit of the coordinate is a function of the coordinate $\hat{x}_s = \mathbf{t}(s, \hat{x})$, the commutators of the digits of the coordinate and of the momentum on the lattice has the form
\be
%       [\hat{x}_s,\hat{p}_r]_{sym}
       [\mathbf{t_{o}}(s,\hat x),\mathbf{t_{o}}(r,\hat p)]
        =    - \Delta p \cdot 3^{-r} \, \sum\limits_{D \in \mathbb{Z}_{3^r/\Delta p}}
        \sum\limits_{\sigma = 1}^{2} \frac{\mathbf{t_{o}}(s, \hat{x}) - \mathbf{t_{o}}(s, \hat{x} - A)}{2\I \sin\left(\pi \Delta p A\right)}
       (-1)^{D+\sigma} %\E^{\pi \I (3D + \sigma)}
       \widehat{T}_{-A},
\ee
\be
%    [\hat{x}_s,\hat{p}_r]_{n.s.} 
    [\mathbf{t_{ns}}(s,\hat x),\mathbf{t_{ns}}(r,\hat p)]
    = - \Delta p \cdot 3^{-r} 
    \sum\limits_{D \in \mathbb{Z}_{3^{r}}/\Delta p} \sum\limits_{\sigma = 1}^{2} 
    \frac{\mathbf{t_{ns}}(s, \hat{x}) - \mathbf{t_{ns}}(s, \hat{x} - A)}{1 - \exp\left(2\pi \I \Delta p A\right)}
    	\widehat{T}_{-A}.
\ee
And for binary systems:
\be
A_2=2^{-r}(D + 1/2),
\ee
\be
[\mathbf{b_{o}}(s,\hat x),\mathbf{b_{o}}(r,\hat p)]
=
\Delta p \, 2^{-r}  \sum\limits_{D \in \mathbb{Z}_{2^{r}/\Delta p}} 
\frac{\mathbf{b_{o}}(s,\hat{x}) - \mathbf{b_{o}}(s, \hat{x} - A_2)}{2 \I \sin (\pi \Delta p\, A_2)}
\widehat{T}_{-A_2},
\ee
\be
%    [\hat{x}_{s}, \hat{p}_{r}] 
[\mathbf{b_{ns}}(s,\hat x),\mathbf{b_{ns}}(r,\hat p)]
= -\Delta p \cdot 2^{-r} 
\sum\limits_{D \in \mathbb{Z}_{2^r/\Delta p}}
\frac{\mathbf{b_{ns}}(s,\hat{x}) - \mathbf{b_{ns}}(s, \hat{x} - A_2)}{1 - \exp({2\pi \I \Delta p A_2})} \widehat{T}_{-A_2}.
\ee

On the line, the commutators of the coordinate and the momentum digits have the form
\begin{equation}
%    [\hat{x}_s,\hat{p}_r]_{sym} 
    [\mathbf{t_{o}}(s,\hat x),\mathbf{t_{o}}(r,\hat p)]
    = - \sum\limits_{D \in \mathbb{Z}} \sum\limits_{\sigma = 1}^{2} 
    \frac{\mathbf{t_{o}}(s, \hat{x})-\mathbf{t_{o}}(s, \hat{x} - 3^{-r}(D + \sigma/3))}{2\pi \I (D + \sigma/3)} 
    (-1)^{D+\sigma}  %\E^{\pi \I (3D + \sigma)} 
    \widehat{T}_{-3^{-r}(D + \sigma/3)},
\end{equation}
\begin{equation}
%    [\hat{x}_s,\hat{p}_r]_{n.s.} 
    [\mathbf{t_{ns}}(s,\hat x),\mathbf{t_{ns}}(r,\hat p)]
    = \sum\limits_{D \in \mathbb{Z}} \sum\limits_{\sigma = 1}^{2}
    \frac{\mathbf{t_{ns}}(s, \hat{x})-\mathbf{t_{ns}}(s, \hat{x} - 3^{-r}(D + \sigma/3))}{2\pi \I (D + \sigma/3)}\widehat{T}_{-3^{-r}(D + \sigma/3)}.
\end{equation}
Similarly, for both binary systems on the real line, we have:
\begin{equation}
    %[\hat{x}_{s}, \hat{p}_{r}]
    [\mathbf{b}(s,\hat{x}),\mathbf{b}(r,\hat{p})] =  
    \sum\limits_{D \in \mathbb{Z}}\frac{\mathbf{b}(s,\hat{x})-\mathbf{b}(s,\hat{x}- 2^{-r}(D + 1/2))}{2\pi \I (D+1/2)}
    \widehat{T}_{-2^{-r}(D+1/2)}.
\end{equation}

In the number sysytem with the base $q$ (in the paper, we consider $q=3$ and $q=2$),
the digit $\hat{x}_s$ in the coordinate representation is given by $c(s,x)$, which has a period $q^{s + 1}$. The shift value $q^{-r}(D + \sigma/q)$ ($\sigma=1,\dots,q-1$) is the period of $c(s,x)$ if
\begin{equation}
    \frac{q^{-r}(D + \sigma/q)}{q^{s + 1}} = q^{-(r + s + 2)}(qD + \sigma) \in \mathbb{Z},\quad.
\end{equation}
i.e., for $-r-s-2 \geqslant 0$. Therefore, $[\hat x_s,\hat p_r] = 0$ (regardless of the system) for $s + r \leqslant -2$. In particular, the fractional part of the momentum commutes with the fractional part of the coordinate, the lowest digit of the momentum does not commute only with the highest digit of the coordinate, and the lowest coordinate digit does not commute only with the highest momentum digit.

\subsection{Coordinate-digit commutator}
Because $\hat{x} = \sum\limits_{s = -n_{-}}^{n_{+}-1}3^{s}\mathbf{t}(s,\hat{x})$, we obtain the commutators of the coordinate and the momentum digit on the lattice:
\begin{equation}
    [\widehat{x}, \mathbf{t_o}(r,\hat p)] = -\Delta p \, 3^{-r} \sum\limits_{D \in \mathbb{Z}_{3^r/\Delta p}} \sum\limits_{\sigma = 1}^{2} 
       % \E^{\pi \I (3D + \sigma)}
    \frac{(-1)^{D+\sigma}A\, \widehat{T}_{-A}}{2\I \sin\left(\pi \Delta p A\right)},
\end{equation}
\begin{equation}
    %[\widehat{x}, \widehat{p}_r]_{n.s.} 
    [\widehat{x}, \mathbf{t_{ns}}(r,\hat p)]
    = - \Delta p  \, 3^{-r}\sum\limits_{D \in \mathbb{Z}_{3^r/\Delta p}} \sum\limits_{\sigma = 1}^{2} 
    \frac{A\,\widehat{T}_{-A}}{1 - \exp(2\pi \I \Delta p\, A)}.
\end{equation}
Similarly for binary systems we have
\be
[\hat x,\mathbf{b_{o}}(r,\hat p)]=
\Delta p \, 2^{-r}  \sum\limits_{D \in \mathbb{Z}_{2^{r}/\Delta p}} 
\frac{A_2\,\widehat{T}_{-A_2}}{2 \I \sin (\pi \Delta p\, A_2)},
\ee
\begin{equation}
    %[\hat{x}, \hat{p}_{r}]_{bin.} 
    [\widehat{x}, \mathbf{b_{ns}}(r,\hat p)]
    =-\Delta p\, 2^{-r} \sum\limits_{D \in \mathbb{Z}_{2^r/\Delta p}}
    \frac{ A_2\, \widehat{T}_{-A_2}}{1 - \exp(2\pi \I \Delta p \, A_2)}.
\end{equation}

On the line, the commutators of the coordinate and the momentum digit have the form
\begin{equation}
    [\widehat{x}, \mathbf{t_o}(r,\hat p)] = \frac{1}{2\pi \I\, 3^r} \sum\limits_{D \in \mathbb{Z}} \sum\limits_{\sigma = 1}^{2} 
    (-1)^{D+\sigma}  %\E^{\pi \I (3D + \sigma)}
    \widehat{T}_{-3^{-r} (D + \sigma/3)},
\end{equation}
\begin{equation}
    [\widehat{x}, \mathbf{t_{ns}}(r,\hat p)] = \frac{1}{2\pi \I\, 3^{r}}\sum\limits_{D \in \mathbb{Z}} \sum\limits_{\sigma = 1}^{2} \, \widehat{T}_{-3^{-r}(D + \sigma/3)}.
\end{equation}
For both binary systems we have
\be
[\hat x,\mathbf{b}(r,\hat p)]=\frac{1}{2\pi \I\,2^r}\sum\limits_{D \in \mathbb{Z}}\widehat{T}_{-2^{-r}(D + 1/2)}.
\ee

\subsection{Coordinate-momentum commutator}
Because $\hat{p} = \sum\limits_{r = -n_{+}}^{n_{-} - 1}3^r \hat{p}_{r}$, we obtain the commutator of the coordinate and momentum on the lattice:
\begin{equation}
    [\widehat{x}, \widehat{p}]_\text{o} = -\Delta p \sum\limits_{r = -n_{+}}^{n_{-} - 1} \sum\limits_{D \in \mathbb{Z}_{3^r/\Delta p}} \sum\limits_{\sigma = 1}^{2}  
    \frac{(-1)^{D+\sigma}A    \,  \widehat{T}_{-A}}{2 \I \sin\left(\pi\,\Delta p\,A\right)},
\end{equation}
\begin{equation}
    [\widehat{x},\widehat{p}]_\text{ns} = -  \Delta p \sum\limits_{r = -n_{+}}^{n_{-} - 1} \sum\limits_{D \in \mathbb{Z}_{3^r/\Delta p}} \sum\limits_{\sigma = 1}^{2} \frac{A\,\widehat{T}_{-A}}{1 - \exp(2\pi \I\, \Delta p A)}.
\end{equation}

In the binary case, we have
\be
[\hat x,\hat p]_\text{bin.o}=
\Delta p  \sum\limits_{r=-n_{+}}^{n_{-} - 1}  \sum\limits_{D \in \mathbb{Z}_{2^{r}/\Delta p}} 
\frac{A_2\,\widehat{T}_{-A_2}}{2 \I \sin (\pi \Delta p\, A_2)},
\ee
\begin{equation}
    [\hat{x}, \hat{p}]_\text{bin.ns} = -\Delta p\sum\limits_{r=-n_{+}}^{n_{-} - 1}\sum\limits_{D \in \mathbb{Z}_{2^r/\Delta p}} \frac{A_2\,\widehat{T}_{-A_2}}{1 - \exp(2\pi \I\, \Delta p \, A_2)}.
\end{equation}
On the line, we obtain the \textit{formal} relations
\begin{equation}\label{[xp]o}
      [\widehat{x}, \widehat{p}]_\text{o} = -\I \hbar \sum\limits_{r \in \mathbb{Z}}\sum\limits_{D \in \mathbb{Z}} \sum\limits_{\sigma = 1}^{2}(-1)^{D + \sigma}  \,  \widehat{T}_{-3^{-r} (D + \sigma/3)},
\end{equation}
\begin{equation}\label{[xp]ns}
    [\widehat{x},\widehat{p}]_\text{ns} = -\I \hbar \sum\limits_{r \in \mathbb{Z}}\sum\limits_{D \in \mathbb{Z}} \sum\limits_{\sigma = 1}^{2} \widehat{T}_{-3^{-r}(D + \sigma/3)},
\end{equation}
\begin{equation}\label{[xp]bin}
    [\hat{x}, \hat{p}]_\text{bin} = -\I \hbar \sum\limits_{r \in \mathbb{Z}}\sum\limits_{D \in \mathbb{Z}}\widehat{T}_{-2^{-r}(D + 1/2)}.
\end{equation}
We recall that $h = 1$ and $\hbar = 1/2\pi$ in the chosen system of units.
\subsection{Renormalization of the commutator on the line}
We obtain a formal decomposition of the commutator in the sum of the shift operators
\eqref{[xp]o}, \eqref{[xp]ns}, \eqref{[xp]bin}.
%\begin{equation}
%      [\widehat{x}, \widehat{p}]_{sym} = -\I \hbar \sum\limits_{r \in \mathbb{Z}}\sum\limits_{D \in \mathbb{Z}} \sum\limits_{\sigma = 1}^{2}(-1)^{D + \sigma}  \,  \widehat{T}_{-3^{-r} (D + \sigma/3)},
%\end{equation}
%\begin{equation}
%    [\widehat{x},\widehat{p}]_{n.s.} = -\I \hbar \sum\limits_{r \in \mathbb{Z}}\sum\limits_{D \in \mathbb{Z}} \sum\limits_{\sigma = 1}^{2} \widehat{T}_{-3^{-r}(D + \sigma/3)}.
%\end{equation}

The values of the shifts for the both ternary systems have the form $-3^{-r}(D + \sigma/3)$, where $r,D \in \mathbb{Z}$, $\sigma \in \{1,2\}$.
 
Let $\mathbb{A}_3$ be a set of numbers whose ternary expansion in a non-symmetric system contains a finite number of nonzero factors with negative powers of three (a finite number of significant ternary digit after the ternary point); $\mathbb{A}_3$ is a group under the summation operation. Then the set of values of the shifts along which the summation occurs has the form $\mathbb{A}_3\setminus\{0\}$. Given that $\widehat{T}_{0} = \hat{1}$, we obtain:
\begin{equation}
    [\hat{x}, \hat{p}]_\text{ns} 
    = -\I \hbar \sum\limits_{A \in \mathbb{A}_3 \setminus \{0\}}\widehat{T}_{A} = -\I \hbar \left(\sum\limits_{A \in \mathbb{A}_3}\widehat{T}_{A} - \hat{1} \right) = \I \hbar \hat{1} - \I \hbar \sum\limits_{A \in \mathbb{A}_3}\widehat{T}_{A}.
\end{equation}
We know that for the particle coordinate and the momentum on the line, there is a canonical commutation relation $[\hat{x}, \hat{p}] = i\hbar \hat{1}$. We, thus, obtain the renormalization
\begin{equation}
    \sum\limits_{A \in \mathbb{A}_3}\widehat{T}_{A} = \sum\limits_{A \in \mathbb{A}_3} \E ^{2\pi \I A \hat{p}} = 0.
\end{equation}
This renormalization is similar to the formal equality $\int\limits_{\mathbb{R}}\E ^{2\pi \I x p}dx = 0$ for all $p \neq 0$ arising in the Fourier transforms.
 
Similarly, for the symmetric system, we can define a set of numbers whose ternary expansion in the symmetric system contains a finite number of nonzero digits after the ternary point $\mathbb{A}_{3^{\prime}}$. To any number $A$ from this set, we can consider a number  $a=A \cdot 3^k ~ (k\in \mathbb{Z}),~ \mathbf{t_o}(0,a) \neq 0,~
\mathbf{t_o}(s,a)=0~\forall s<0$. Then for the symmetric system, we get the following renormalization: 
\begin{equation}
    \sum\limits_{A \in \mathbb{A}_{3^{\prime}}} \widehat{T}_{A}(-1)^{a} = \sum\limits_{A \in \mathbb{A}_{3^{\prime}}} \E^{2\pi \I A \hat{p}} (-1)^{a} = 0.
\end{equation}

In the binary case, for both systems, we have the following renormalization \cite{bin-ivanov}
\begin{equation}
    \sum\limits_{A \in \mathbb{A}_2}\widehat{T}_{A} = \sum\limits_{A \in \mathbb{A}_2} \E ^{2\pi \I A \hat{p}} = 0,
\end{equation}
where $\mathbb{A}_2$ is a set of numbers whose binary expansion in a non-symmetric system contains a finite number of nonzero binary digits after the binary point; $\mathbb{A}_2$ is a group under the summation operation.

\section{Conclusion}
We have generalized the results of the paper \cite{bin-ivanov} to a binary symmetric number system and two ternary number systems.

The binary and ternary expansions of the coordinate and momentum operators that we constructed here are applicable not only in the field of quantum computations but also in numerically solving partial differential equations on a classical computer. 
The natural appearance of renormalizations is particularly interesting.
Renormalizing infinite and finite quantities (on the lattice) allows finding the renormalization numerically by passing from the lattice to the limit of a continuous quantity. 
Time and energy can also be considered as a coordinate and a momentum \cite{dinamic-t}, which allows applying the same renormalization methods to them. 
Renormalizations in this context are probably related to the quantum theory of measurements
(see \cite{q-ivanov}, \cite{pr-ivanov} and the references therein for the quantum theory of measurements). The representation of the
coordinate and the momentum in the form of a binary or ternary expansion assumes that the coordinate and the momentum are not observed in the experiment. 
Instead, individual digits of the coordinate and the momentum are directly observed. 
The measurement of a binary digit of the spatial coordinate corresponds to the
passage/nonpassage of a particle through a diffraction grating.

\subsection*{Acknowledgments}
The authors thank the participants of the Conference Phystech-Quant 2020 (Moscow Institute of Physics and Technology, 2020),
of the seminar ``Quantum Physics and Quantum Information'' (Moscow Institute of Physics and Technology, 2019), 
and of the section of theoretical physics of the 62-nd Science Conference of MIPT (Moscow Institute of Physics and Technology, 2019) for the discussions. 
Separately, the authors thank I. V. Volovich, Yu. M. Belosov, E. I. Zelenov, V. I. Manko, I. V. Maresin, N. N. Nepoyvoda, D. A. Podlessnykh, D. V. Prokhorenko , V. Zh. Sakbayev, M. V. Suslov, A. S. Trushechkin, L. Y. Fedichkin, S. N. Filippov, N. N. Shamarov, and other colleagues with whom the material of this paper was discussed.

\appendix
\section{Appendices}

\subsection{Rectangular function on the lattice}
Consider a function $c$, which is a periodical rectangular function with a period $\tau$ ($T$ nodes of lattice), i.e., which is equal to 0 on the whole period except nodes $[s,f]$. In contrast, in nodes $s, f$, and every node between them the function is equal to  $A$. For $k \neq 0$: 
$$
\widetilde{c}_{k} = \frac{1}{T\Delta x}\sum\limits_{\alpha = s}^{f}A \E^{\frac{2 \pi \I k \alpha}{T}} \Delta x = \frac{A}{T}\sum\limits_{\alpha = s}^{f}\E^{\frac{2 \pi \I k \alpha}{T}},
$$
the sum of this series can be calculated as a sum of geometric progression:
$$
\frac{A}{T}\sum\limits_{\alpha = s}^{f}\E^{\frac{2 \pi \I k \alpha}{T}} = \frac{A}{T} \E^{\frac{2 \pi \I k s}{T}} \left( \frac{1 - \E^{\frac{2 \pi \I k (f - s + 1)}{T}}}{1 - \E^{\frac{2\pi \I k}{T}}}\right).
$$
Then:
$$
    c^\alpha = \sum\limits_{k \in \Delta x \cdot \mathbb{Z}_N} \frac{A}{T} \E^{\frac{2 \pi \I k s}{T}} \left( \frac{1 - \E^{\frac{2 \pi \I k (f - s + 1)}{T}}}{1 - \E^{\frac{2\pi \I k}{T}}}\right) \E^{-\frac{2 \I \pi k \alpha}{T}}
$$

\subsection{Calculating digits of the coordinate and the momentum}
We calculate the Fourier amplitudes of the momentum digits explicitly:
$$
\tilde{\mathbf{t}}_{n_{+}n_{-}}(r,A) = \frac{1}{N}\sum\limits_{p \in \Delta p \cdot \mathbb{Z}_{N}}\E^{-2\pi \I p A}\mathbf{t}_{n_{+}n_{-}}(r,p).
$$
As the digit of the coordinate is a periodical partially-constant function of the coordinate, it can be expanded in a sum of periodic rectangular functions:
$$
\mathbf{t_{o}}= f_{+1} + f_{-1},
$$
$$
\mathbf{t_{ns}}= f_{1} + f_{2},
$$
where index $i$ is the amplitude of the periodic rectangular function  $f_{i}$.
Using this formula, we can calculate the Fourier transform for both systems explicitly:
$$
\mathbf{t_{ns}}(r,\hat p)= \widehat{1} +   \left (\sum\limits_{D \in \mathbb{Z}_{3^r/\Delta p}} \sum\limits_{\sigma = 1}^{2} 3^{-(n_{+} + r + 1)} (\E^{\frac{2\pi \I \sigma}{3}} +  2\E^{\frac{4\pi \I \sigma}{3}}) \frac{1 - \E^{\frac{2\pi \I \sigma}{3}}}{1 - \E^{\frac{2\pi \I (3D + \sigma)}{3^{n_{+} + r + 1}}}} \widehat{T}_{-3^{-(r + 1)}(3D + \sigma)}\right).
$$
%Note that 
$$
(\E^{\frac{2 \pi \I \sigma}{3}} + 2\E^{\frac{4\pi \sigma}{3}})(1 - \E^{\frac{2\pi \I \sigma}{3}}) = 2\left(\cos\left(\frac{2\pi\sigma}{3}\right)\right) - 2 = -3,
$$
Hence,
$$
    \mathbf{t_{ns}}(r,\hat p) = \widehat{1} -\sum\limits_{D \in \mathbb{Z}_{3^{r}}/\Delta p} \sum\limits_{\sigma = 1}^{2} 3^{-(n_{+} + r)}\frac{\widehat{T}_{-3^{-(r + 1)}(3D + \sigma)}}{1 - \exp\left(\frac{2\pi \I (D + \sigma /3)}{3^{n_{+} + r}}\right)}.
$$

Similarly, we can calculate the Fourier transform for the symmetric system:
$$
\mathbf{t_{o}}(r,\hat p) = \sum\limits_{D \in \mathbb{Z}_{3^r/\Delta p}} \sum\limits_{\sigma = 1}^{2} 3^{-(n_+ + r)} \frac{\E^{\pi \I (3D + \sigma)}}{2\I \sin\left(\frac{\pi(3D + \sigma)}{3^{n_+ + r + 1}}\right)}\widehat{T}_{-3^{-(r)} (D + \sigma/3)}.
$$
Similarly, after replacing $n_{+} \leftrightarrow n_{-}$, $\Delta x \leftrightarrow \Delta p$, $r \to s$ and $A \to B$, we obtain the operator of the digit of the coordinate in terms of the shifts in the momentum:
$$
\hat{x}_s = \mathbf{t}_{n_{-}n_{+}}(s, \hat{x}),
$$
Then, for particular systems, we obtain the following expressions:
$$
 \mathbf{t_{ns}}(s,\hat x)
% \hat{x}_{s_{n.s.}} 
 = \widehat{1} -\sum\limits_{D \in \mathbb{Z}_{3^{s}/\Delta x}} \sum\limits_{\sigma = 1}^{2} 3^{-(n_{-} + s)}\frac{\widehat{S}_{3^{-(s + 1)}(3D + \sigma)}}{1 - \exp\left(\frac{2\pi \I (D + \sigma /3)}{3^{n_{-} + s}}\right)},
$$
$$
 \mathbf{t_{o}}(s,\hat x)
%\hat{x}_{s_{sym}} 
= \sum\limits_{D \in \mathbb{Z}_{3^s/\Delta x}} \sum\limits_{\sigma = 1}^{2} 3^{-(n_- + s)} \frac{\E^{\pi \I (3D + \sigma)}}{2\I \sin\left(\frac{\pi(3D + \sigma)}{3^{n_- + s + 1}}\right)}\widehat{S}_{3^{-(s)} (D + \sigma/3)}.
$$

\subsection{Digits on the lattice for the ternary system}
\subsubsection{Symmetric ternary system}

The digits for the symmetric ternary system on the finite lattice with $n_{-} = 1$ can be set as:
\begin{equation}
\textbf{t}_{\textbf{o}}(0,x) = 
\left\{
\begin{array}{cc}
	0,& x \in \{-1/3,0,1/3\} \; \text{(mod~3)},\\
	1,&  x \in \{2/3,1,4/2 \} \; \text{(mod~3)},\\
	-1,& x \in \{5/3,2,7/3 \} \; \text{(mod~3)}.\\
\end{array}\right..
\end{equation}

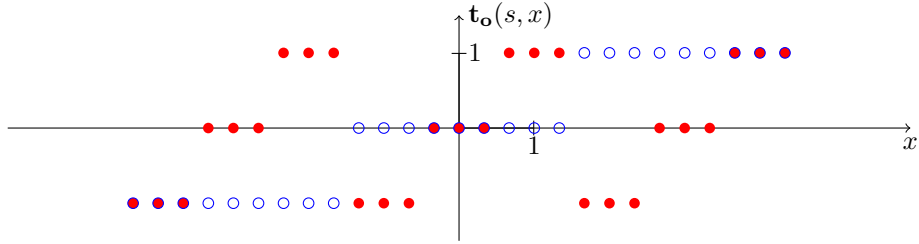
\begin{figure}[h]
    \centering
\begin{tikzpicture}
\draw [->] (-6,0) -- (6,0) node [below] {$x$};
\draw [-|] (0,0) -- (1,0) node [below] {1};
\draw [-|] (0,0) -- (0,1) node [right] {1};
\draw [->] (0,-1.5) -- (0,1.5) node [right] {$\mathbf{t_o}(s,x)$};

\fill[red] (-4.333,-1) circle (2pt);
\fill[red] (-4,-1) circle (2pt);
\fill[red] (-3.666,-1) circle (2pt);

\fill[red] (-3.333,0) circle (2pt);
\fill[red] (-3,0) circle (2pt);
\fill[red] (-2.666,0) circle (2pt);

\fill[red] (-2.333,1) circle (2pt);
\fill[red] (-2,1) circle (2pt);
\fill[red] (-1.666,1) circle (2pt);

\fill[red] (-0.666,-1) circle (2pt);
\fill[red] (-1,-1) circle (2pt);
\fill[red] (-1.333,-1) circle (2pt);

\fill[red] (-0.333,0) circle (2pt);
\fill[red] (0,0) circle (2pt);
\fill[red] (0.333,0) circle (2pt);

\fill[red] (0.666,1) circle (2pt);
\fill[red] (1,1) circle (2pt);
\fill[red] (1.333,1) circle (2pt);

\fill[red] (1.666,-1) circle (2pt);
\fill[red] (2,-1) circle (2pt);
\fill[red] (2.333,-1) circle (2pt);

\fill[red] (2.666,0) circle (2pt);
\fill[red] (3,0) circle (2pt);
\fill[red] (3.333,0) circle (2pt);

\fill[red] (3.666,1) circle (2pt);
\fill[red] (4,1) circle (2pt);
\fill[red] (4.333,1) circle (2pt);

\draw [color=blue] (-1.666,-1) circle (2pt);
\draw [color=blue] (-2,-1) circle (2pt);
\draw [color=blue] (-2.333,-1) circle (2pt);
\draw [color=blue] (-2.666,-1) circle (2pt);
\draw [color=blue] (-3,-1) circle (2pt);
\draw [color=blue] (-3.333,-1) circle (2pt);
\draw [color=blue] (-3.666,-1) circle (2pt);
\draw [color=blue] (-4,-1) circle (2pt);
\draw [color=blue] (-4.333,-1) circle (2pt);

\draw [color=blue] (-1.333,0) circle (2pt);
\draw [color=blue] (-1,0) circle (2pt);
\draw [color=blue] (-0.666,0) circle (2pt);
\draw [color=blue] (-0.333,0) circle (2pt);
\draw [color=blue] (0,0) circle (2pt);
\draw [color=blue] (0.333,0) circle (2pt);
\draw [color=blue] (0.666,0) circle (2pt);
\draw [color=blue] (1,0) circle (2pt);
\draw [color=blue] (1.333,0) circle (2pt);

\draw [color=blue] (1.666,1) circle (2pt);
\draw [color=blue] (2,1) circle (2pt);
\draw [color=blue] (2.333,1) circle (2pt);
\draw [color=blue] (2.666,1) circle (2pt);
\draw [color=blue] (3,1) circle (2pt);
\draw [color=blue] (3.333,1) circle (2pt);
\draw [color=blue] (3.666,1) circle (2pt);
\draw [color=blue] (4,1) circle (2pt);
\draw [color=blue] (4.333,1) circle (2pt);

\end{tikzpicture}
\caption{Plot of the value of the ternary digit number $s$ for a finite lattice ($n_{-} = 1$), $s=0$ -- red filled circles, $s=1$ -- blue circles.}
\end{figure}

For arbitrary $n_{-}$, we can use the following expression:

\begin{equation}
\textbf{t}_{\textbf{o}}(s,x) = 
\begin{cases}
	0,\, x \in \{-\frac{3^{s}}{2} + \frac{3^{-n_{-}}}{2}, \frac{1}{2} - 3 \cdot  \frac{3^{-n_{-}}}{2}\, ... \, , 0, ... \, , \frac{3^{s}}{2} - \frac{3^{-n_{-}}}{2} \} \; \text{(mod~} 3^{s+1}),\\
	1, \, x \in \{\frac{3^{s}}{2} + \frac{3^{-n_{-}}}{2}, ... \, , 1, ... \, , \frac{3^{s + 1}}{2} - \frac{3^{-n_{-}}}{2} \} \; \text{(mod~} 3^{s+1}),\\
	-1, \, x \in \{\frac{3^{s + 1}}{2} + \frac{3^{-n_{-}}}{2},...\,,2, ... \, ,\frac{5 \cdot 3^{s}}{2} - \frac{3^{-n_{-}}}{2} \} \; \text{(mod~} 3^{s+1}).\\
\end{cases}
\end{equation}
\textit{Remark:} Step between nodes is $\Delta x = 3^{-n_{-}}$.

\subsubsection{Non-symmetric ternary system}
The non-symmetric ternary system can be obtained from a symmetric one by translation along the $x$-axis for $\left[\frac{3^{n_{-}}}{2} \right] + 3^{n_{-}}$ nodes, which is $\left( \frac{3^{n_{-}} - 1}{2} + 3^{n_{-}}\right) \cdot 3^{-n_{-}} \cdot 3^{s} = \frac{3}{2} - \frac{3^{s - n_{-}}}{2}$ and translation among $y$-axis for 1. Therefore, for $n_{-} = 1$ we obtain the following expression:
\begin{equation}
\textbf{t}_{\textbf{ns}}(0,x) = 
\begin{cases}
	0,\, x \in \{0,1/3,2/3\} \; \text{(mod~3)},\\
	1, \, x \in \{1,4/3,5/3 \} \; \text{(mod~3)},\\
	-1, \, x \in \{2,7/3,8/3 \} \; \text{(mod~3)}.\\
\end{cases}
\end{equation}

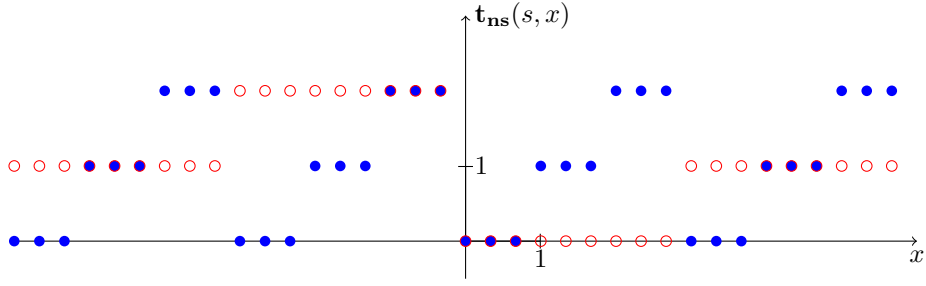
\begin{figure}[h!]
    \centering
\begin{tikzpicture}
\draw [->] (-6,0) -- (6,0) node [below] {$x$};
\draw [-|] (0,0) -- (1,0) node [below] {1};
\draw [-|] (0,0) -- (0,1) node [right] {1};
\draw [->] (0,-0.5) -- (0,3) node [right] {$\mathbf{t_{ns}}(s,x)$};

\fill[blue] (-6,0) circle (2pt);
\fill[blue] (-5.666,0) circle (2pt);
\fill[blue] (-5.333,0) circle (2pt);

\fill[blue] (-5,1) circle (2pt);
\fill[blue] (-4.666,1) circle (2pt);
\fill[blue] (-4.333,1) circle (2pt);

\fill[blue] (-4,2) circle (2pt);
\fill[blue] (-3.666,2) circle (2pt);
\fill[blue] (-3.333,2) circle (2pt);

\fill[blue] (-3,0) circle (2pt);
\fill[blue] (-2.666,0) circle (2pt);
\fill[blue] (-2.333,0) circle (2pt);

\fill[blue] (-2,1) circle (2pt);
\fill[blue] (-1.666,1) circle (2pt);
\fill[blue] (-1.333,1) circle (2pt);

\fill[blue] (-1,2) circle (2pt);
\fill[blue] (-0.666,2) circle (2pt);
\fill[blue] (-0.333,2) circle (2pt);

\fill[blue] (0,0) circle (2pt);
\fill[blue] (0.333,0) circle (2pt);
\fill[blue] (0.666,0) circle (2pt);

\fill[blue] (1,1) circle (2pt);
\fill[blue] (1.333,1) circle (2pt);
\fill[blue] (1.666,1) circle (2pt);

\fill[blue] (2,2) circle (2pt);
\fill[blue] (2.333,2) circle (2pt);
\fill[blue] (2.666,2) circle (2pt);

\fill[blue] (3,0) circle (2pt);
\fill[blue] (3.333,0) circle (2pt);
\fill[blue] (3.666,0) circle (2pt);

\fill[blue] (4,1) circle (2pt);
\fill[blue] (4.333,1) circle (2pt);
\fill[blue] (4.666,1) circle (2pt);

\fill[blue] (5,2) circle (2pt);
\fill[blue] (5.333,2) circle (2pt);
\fill[blue] (5.666,2) circle (2pt);

\draw[red] (0,0) circle (2pt);
\draw[red] (0.333,0) circle (2pt);
\draw[red] (0.666,0) circle (2pt);
\draw[red] (1,0) circle (2pt);
\draw[red] (1.333,0) circle (2pt);
\draw[red] (1.666,0) circle (2pt);
\draw[red] (2,0) circle (2pt);
\draw[red] (2.333,0) circle (2pt);
\draw[red] (2.666,0) circle (2pt);

\draw[red] (3,1) circle (2pt);
\draw[red] (3.333,1) circle (2pt);
\draw[red] (3.666,1) circle (2pt);
\draw[red] (4,1) circle (2pt);
\draw[red] (4.333,1) circle (2pt);
\draw[red] (4.666,1) circle (2pt);
\draw[red] (5,1) circle (2pt);
\draw[red] (5.333,1) circle (2pt);
\draw[red] (5.666,1) circle (2pt);

\draw[red] (-0.333,2) circle (2pt);
\draw[red] (-0.666,2) circle (2pt);
\draw[red] (-1,2) circle (2pt);
\draw[red] (-1.333,2) circle (2pt);
\draw[red] (-1.666,2) circle (2pt);
\draw[red] (-2,2) circle (2pt);
\draw[red] (-2.333,2) circle (2pt);
\draw[red] (-2.666,2) circle (2pt);
\draw[red] (-3,2) circle (2pt);

\draw[red] (-6,1) circle (2pt);
\draw[red] (-3.333,1) circle (2pt);
\draw[red] (-3.666,1) circle (2pt);
\draw[red] (-4,1) circle (2pt);
\draw[red] (-4.333,1) circle (2pt);
\draw[red] (-4.666,1) circle (2pt);
\draw[red] (-5,1) circle (2pt);
\draw[red] (-5.333,1) circle (2pt);
\draw[red] (-5.666,1) circle (2pt);

\end{tikzpicture}
    \caption{Plot of the value of the ternary digit number $s$ for a finite lattice ($n_{-} = 1$), $s=0$ -- blue filled circles, $s=1$ -- red circles.}
\end{figure}

In full analogy with the symmetric system we have the following formula for arbitrary $n_{-}$:
\begin{equation}
	\textbf{t}_{\textbf{ns}}(s,x) = 
\begin{cases}
	0,\, x \in \{0,3^{-n_{-}}, ... \, , 3^{s} - 3^{-n_{-}}\} \; \text{(mod~3}^{s+1}),\\
	1, \, x \in \{3^{s},3^{-n_{-}}, ... \, , 2 \cdot 3^{s} - 3^{-n_{-}} \} \; \text{(mod~3}^{s+1}),\\
	-1, \, x \in \{2 \cdot 3^{s},3^{-n_{-}}, ... \, , 3^{s+1} - 3^{-n_{-}} \} \;  \text{(mod~3}^{s+1}).\\
\end{cases}
\end{equation}

\subsection{Digits on the lattice for the binary system}
\subsubsection{Non-symmetric binary system}
For the non-symmetric binary system, we can write the following expression:
\begin{equation}
\textbf{b}_{\textbf{ns}}(s,x) = 
\begin{cases}
0,  x \in \{0, ... \, , 2^{s} - 2^{-n_{-}} \} \; \text{(mod~2}^{s+1}) \\
1, x \in \{2^{s} , ... \, , 2^{s+1} - 2^{-n_{-}} \} \; \text{(mod~2}^{s+1}) \\
\end{cases}
\end{equation}

\begin{figure}[h]
\centering
\begin{tikzpicture}
\draw [->] (-5,0) -- (5,0) node [below] {$x$};
\draw [-|] (0,0) -- (1,0) node [below] {1};
\draw [-|] (0,0) -- (0,1) node [right] {1};
\draw [->] (0,-2) -- (0,2) node [right] {$\mathbf{b_{ns}}(s,x)$};

\fill[green] (-4,0) circle (2pt);
\fill[green] (-3.5,0) circle (2pt);

\fill[green] (-3,1) circle (2pt);
\fill[green] (-2.5,1) circle (2pt);

\fill[green] (-2,0) circle (2pt);
\fill[green] (-1.5,0) circle (2pt);

\fill[green] (-1,1) circle (2pt);
\fill[green] (-0.5,1) circle (2pt);

\fill[green] (0,0) circle (2pt);
\fill[green] (0.5,0) circle (2pt);

\fill[green] (1,1) circle (2pt);
\fill[green] (1.5,1) circle (2pt);

\fill[green] (2,0) circle (2pt);
\fill[green] (2.5,0) circle (2pt);

\fill[green] (3,1) circle (2pt);
\fill[green] (3.5,1) circle (2pt);

\draw[blue] (0,0) circle (2pt);
\draw[blue] (0.5,0) circle (2pt);
\draw[blue] (1,0) circle (2pt);
\draw[blue] (1.5,0) circle (2pt);

\draw[blue] (2,1) circle (2pt);
\draw[blue] (2.5,1) circle (2pt);
\draw[blue] (3,1) circle (2pt);
\draw[blue] (3.5,1) circle (2pt);

\draw[blue] (-0.5,1) circle (2pt);
\draw[blue] (-1,1) circle (2pt);
\draw[blue] (-1.5,1) circle (2pt);
\draw[blue] (-2,1) circle (2pt);

\draw[blue] (-2.5,0) circle (2pt);
\draw[blue] (-3,0) circle (2pt);
\draw[blue] (-3.5,0) circle (2pt);
\draw[blue] (-4,0) circle (2pt);

\draw (0,0) node[cross=3pt,black]{};
\draw (0.5,1) node[cross=3pt,black]{};
\draw (1,0) node[cross=3pt,black]{};
\draw (1.5,1) node[cross=3pt,black]{};
\draw (2,0) node[cross=3pt,black]{};
\draw (2.5,1) node[cross=3pt,black]{};
\draw (3,0) node[cross=3pt,black]{};
\draw (3.5,1) node[cross=3pt,black]{};
\draw (-0.5,1) node[cross=3pt,black]{};
\draw (-1,0) node[cross=3pt,black]{};
\draw (-1.5,1) node[cross=3pt,black]{};
\draw (-2,0) node[cross=3pt,black]{};
\draw (-2.5,1) node[cross=3pt,black]{};
\draw (-3,0) node[cross=3pt,black]{};
\draw (-3.5,1) node[cross=3pt,black]{};
\draw (-4,0) node[cross=3pt,black]{};

\end{tikzpicture}
\caption{Plot of the value of the binary digit number $s$ on the lattice for an ``non-symmetric system,'' ($n_{-} = 1$), $s = 0$ -- green filled circles, $s = 1$ -- blue circles, $s=-1$ -- crosses.\label{bin-graph-ns}}
\end{figure}
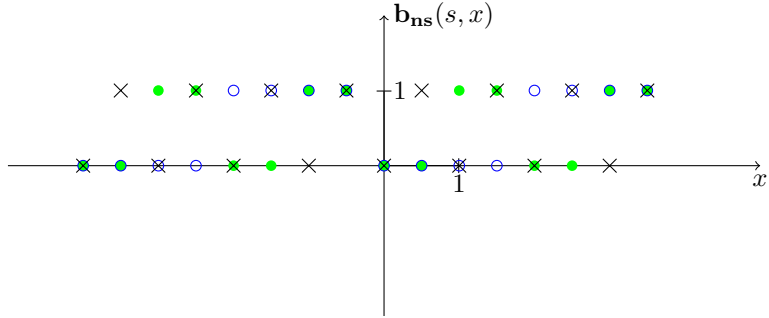

\textit{Remark:} Step between nodes is $\Delta x = 2^{-n_{-}}$.

\subsubsection{Symmetric binary system}
The binary symmetric system can be obtained from a non-symmetric one by a shift on $2^{-n_{-}}$ along $x$-axis and on $-1/2$ along the $y$-axis:
\begin{equation}
\textbf{b}_{\textbf{0}}(s,x) = 
\begin{cases}
-1/2,  x \in \{2^{-n_{-}-1}, ... \, , 2^{s} - 2^{-n_{-} - 1} \} \; \text{(mod~2}^{s+1}) \\
1/2, x \in \{2^{s} + 2^{-n_{-} - 1}, ... \, , 2^{s+1} - 2^{-n_{-} - 1} \} \; \text{(mod~2}^{s+1}) \\
\end{cases}
\end{equation}
\begin{figure}[h]
\centering
\begin{tikzpicture}
\draw [->] (-5,0) -- (5,0) node [below] {$x$};
\draw [-|] (0,0) -- (1,0) node [below] {1};
\draw [-|] (0,0) -- (0,1) node [right] {1};
\draw [->] (0,-2) -- (0,2) node [right] {$\mathbf{b_{o}}(s,x)$};

\fill[red] (-3.75,-0.5) circle (2pt);
\fill[red] (-3.25,-0.5) circle (2pt);

\fill[red] (-2.75,0.5) circle (2pt);
\fill[red] (-2.25,0.5) circle (2pt);

\fill[red] (-1.75,-0.5) circle (2pt);
\fill[red] (-1.25,-0.5) circle (2pt);

\fill[red] (-0.75,0.5) circle (2pt);
\fill[red] (-0.25,0.5) circle (2pt);

\fill[red] (0.25,-0.5) circle (2pt);
\fill[red] (0.75,-0.5) circle (2pt);

\fill[red] (1.25,0.5) circle (2pt);
\fill[red] (1.75,0.5) circle (2pt);

\fill[red] (2.25,-0.5) circle (2pt);
\fill[red] (2.75,-0.5) circle (2pt);

\fill[red] (3.25,0.5) circle (2pt);
\fill[red] (3.75,0.5) circle (2pt);

\draw[blue] (0.25, -0.5) circle (2pt);
\draw[blue] (0.75, -0.5) circle (2pt);
\draw[blue] (1.25, -0.5) circle (2pt);
\draw[blue] (1.75, -0.5) circle (2pt);

\draw[blue] (2.25, 0.5) circle (2pt);
\draw[blue] (2.75, 0.5) circle (2pt);
\draw[blue] (3.25, 0.5) circle (2pt);
\draw[blue] (3.75, 0.5) circle (2pt);

\draw[blue] (-0.25, 0.5) circle (2pt);
\draw[blue] (-0.75, 0.5) circle (2pt);
\draw[blue] (-1.25, 0.5) circle (2pt);
\draw[blue] (-1.75, 0.5) circle (2pt);

\draw[blue] (-2.25, -0.5) circle (2pt);
\draw[blue] (-2.75, -0.5) circle (2pt);
\draw[blue] (-3.25, -0.5) circle (2pt);
\draw[blue] (-3.75, -0.5) circle (2pt);

\draw (0+0.25,-0.5) node[cross=3pt,black]{};
\draw (0.5+0.25,0.5) node[cross=3pt,black]{};
\draw (1+0.25,-0.5) node[cross=3pt,black]{};
\draw (1.5+0.25,0.5) node[cross=3pt,black]{};
\draw (2+0.25,-0.5) node[cross=3pt,black]{};
\draw (2.5+0.25,0.5) node[cross=3pt,black]{};
\draw (3+0.25,-0.5) node[cross=3pt,black]{};
\draw (3.5+0.25,0.5) node[cross=3pt,black]{};
\draw (-0.5+0.25,0.5) node[cross=3pt,black]{};
\draw (-1+0.25,-0.5) node[cross=3pt,black]{};
\draw (-1.5+0.25,0.5) node[cross=3pt,black]{};
\draw (-2+0.25,-0.5) node[cross=3pt,black]{};
\draw (-2.5+0.25,0.5) node[cross=3pt,black]{};
\draw (-3+0.25,-0.5) node[cross=3pt,black]{};
\draw (-3.5+0.25,0.5) node[cross=3pt,black]{};
\draw (-4+0.25,-0.5) node[cross=3pt,black]{};

\end{tikzpicture}
\caption{Plot of the value of the binary digit number $s$ on the lattice for a ``symmetric system,'' ($n_{-} = 1$), $s = 0$ -- red filled circles, $s = 1$ -- blue circles, $s=-1$ -- black crosses.}
\end{figure}
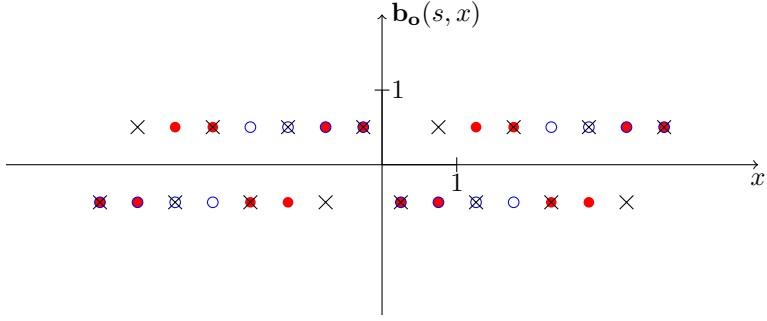

\subsection{Checking the integral ternary representation for $x > 0$}
We prove the validity of integral representation for $x > 0$ by the direct calculation. We have
$$
\mathbf{t_{ns}}(s,x) = 
\begin{cases}
  0,& s > \log_3 x\\
  1,& s \in (\log_3 \frac{x}{2}, \log_3 x]\\
  2,& s \in (\log_3\frac{x}{3}, \log_3\frac{x}{2}]\\
  \vdots& \vdots\\
  0,& s \in (\log_3\frac{x}{3k - 2}, \log_3\frac{x}{3k - 3}]\\
  1,& s \in (\log_3\frac{x}{3k - 1}, \log_3\frac{x}{3k - 2}]\\
  2,& s \in (\log_3\frac{x}{3k}, \log_3\frac{x}{3k - 1}]\\
  \vdots& \vdots\\
\end{cases}
$$
Hence,
$$
\int\limits_{-\infty}^{\log_3 x}\mathbf{t_{ns}}(s,x)3^sds = \sum\limits_{k = 1}^{\infty}\left(\int\limits_{\log_3 (\frac{x}{3k -1})}^{\log_3 (\frac{x}{3k -2})}\E^{s \ln 3}ds + 2 \cdot \int\limits_{\log_3 (\frac{x}{3k})}^{\log_3 (\frac{x}{3k -1})}\E^{s \ln 3}ds \right) =
$$
$$
= \sum\limits_{k = 1}^{\infty}\left(\frac{1}{\ln 3}3^s\bigg|_{\log_3 (\frac{x}{3k -1})}^{\log_3 (\frac{x}{3k -2})} + \frac{2}{\ln 3}3^s\bigg|_{\log_3 (\frac{x}{3k})}^{\log_3 (\frac{x}{3k - 1})} \right) 
= \frac{x}{\ln 3} \underbrace{\sum\limits_{k = 1}^{\infty}\left(\frac{1}{3k - 2} + \frac{1}{3k - 1} - \frac{2}{3k}  \right)}_{\ln 3} = x
$$

\subsection{Examples}
Everywhere in this section, $\Delta x = 1, \, x \in \mathbb{Z}_{N} \in \{0,1, ... ,N - 1  \}$, the coordinates and the momenta are numbered by ternary numbers, which are marked with a lower index 3, and $\Delta p = 3^{-n} = 1/N$.

\subsubsection{The case $n = 1$ and $N = 3^1 = 3$. Symmetric system} In this case, we have 
$$\hat x = \hat x_0 = \left(
\begin{array}{ccc}
     +1&0&0  \\
     0&0&0  \\
     0&0&-1
\end{array}
\right) = \hat{s}_{z}
$$

$$
\hat p_{-1} =\frac{1}{\sqrt{3}} \left(
\begin{array}{ccc}
    0 & \I & -\I \\
    -\I & 0 & \I \\
    \I & -\I & 0
\end{array}
\right) = \frac{1}{\sqrt{3}} \left(\sqrt{2}\hat{s}_{y} + 2\hat{s}_{y}\hat{s}_{x} + \I \hat{s}_{z} \right)
$$%тут надо поискать выражение через степени не выше 2; кажется, получилось!

\subsubsection{The case $n = 1$ and $N = 3^1 = 3$. Non-symmetric system}
Here, we obtain the following results::
$$\hat x = \hat x_0 = \left(
\begin{array}{ccc}
     2&0&0  \\
     0&1&0  \\
     0&0&0
\end{array}
\right) = \hat{1} + \hat{s}_z; \quad
\hat p_{-1} =\frac{1}{6} \left(
\begin{array}{ccc}
     6& 1 - \sqrt{3}\,\I& 1 + \sqrt{3}\,\I  \\
     1 + \sqrt{3}\,\I& 6 & 1 - \sqrt{3}\,\I \\
     1 - \sqrt{3}\,\I& 1 + \sqrt{3}\,\I& 6
\end{array}
\right) =
$$
$$=\hat{1} +  \frac{\sqrt{2}}{6}\hat{s}_{x} + \frac{1}{\sqrt{6}}\hat{s}_{y} - \frac{\sqrt{3}}{6}(2\hat{s}_{y}\hat{s}_{x} + \I \hat{s}_{z})
$$

\subsubsection{The case $n = 2$ and $N = 3^2 = 9$. Non - symmetric system}
%В больших матрицах я всё-таки не уверен, хотя, кажется, на этот раз сошлось.
$$\hat x =\hat x_0 + 3\cdot\hat x_1=\text{diag}(8;7;6;5;4;3;2;1;0),$$
%=  \left(
%    \begin{array}{ccccccccc}
%         8& 0 &0 &0 &0 &0 &0 &0 &0\\
%         0& 7 &0 &0 &0 &0 &0 &0 &0\\
%         0& 0 &6 &0 &0 &0 &0 &0 &0\\
%         0& 0 &0 &5 &0 &0 &0 &0 &0\\
%         0& 0 &0 &0 &4 &0 &0 &0 &0\\
%         0& 0 &0 &0 &0 &3 &0 &0 &0\\
%         0& 0 &0 &0 &0 &0 &2 &0 &0\\
%         0& 0 &0 &0 &0 &0 &0 &1 &0\\
%         0& 0 &0 &0 &0 &0 &0 &0 &0
%    \end{array}
%    \right)
%    \vspace{10truemm}$;
$$\hat x_0=\text{diag}(2;1;0;2;1;0;2;1;0),$$
%= \left(
%    \begin{array}{ccccccccc}
%         2& 0 &0 &0 &0 &0 &0 &0 &0\\
%         0& 1 &0 &0 &0 &0 &0 &0 &0\\
%         0& 0 &0 &0 &0 &0 &0 &0 &0\\
%         0& 0 &0 &2 &0 &0 &0 &0 &0\\
%         0& 0 &0 &0 &1 &0 &0 &0 &0\\
%         0& 0 &0 &0 &0 &0 &0 &0 &0\\
%         0& 0 &0 &0 &0 &0 &2 &0 &0\\
%         0& 0 &0 &0 &0 &0 &0 &1 &0\\
%         0& 0 &0 &0 &0 &0 &0 &0 &0
%    \end{array}
%    \right)$,
$$\hat x_1=\text{diag}(2;2;2;1;1;1;0;0;0).$$
% = \left(
%    \begin{array}{ccccccccc}
%         2& 0 &0 &0 &0 &0 &0 &0 &0\\
%         0& 2 &0 &0 &0 &0 &0 &0 &0\\
%         0& 0 &2 &0 &0 &0 &0 &0 &0\\
%         0& 0 &0 &1 &0 &0 &0 &0 &0\\
%         0& 0 &0 &0 &1 &0 &0 &0 &0\\
%         0& 0 &0 &0 &0 &1 &0 &0 &0\\
%         0& 0 &0 &0 &0 &0 &0 &0 &0\\
%         0& 0 &0 &0 &0 &0 &0 &0 &0\\
%         0& 0 &0 &0 &0 &0 &0 &0 &0
%    \end{array}
%    \right)$.
 
Let's denote
$$
    E_{n} =  \frac{1}{\exp\left(\frac{-2 \pi \I n}{9}\right) - 1},
    $$
    Then
$$
\hat{p}_{-1} = \frac{1}{3}\left(
\begin{array}{ccccccccc}
     3&E_{8} &E_{7} & 0& E_{5}& E_{4}&0 &E_{2} & E_{1}\\
     E_{1} & 3 &E_{8} &E_{7} & 0& E_{5}& E_{4}&0 &E_{2}\\
     E_{2} &E_{1} & 3 &E_{8} &E_{7} & 0& E_{5}& E_{4}&0\\
     0&E_{2} &E_{1} & 3 &E_{8} &E_{7} & 0& E_{5}& E_{4}\\
     E_{4} &0&E_{2} &E_{1} & 3 &E_{8} &E_{7} & 0& E_{5}\\
     E_{5} &E_{4} &0&E_{2} &E_{1} & 3 &E_{8} &E_{7} & 0\\
     0 &E_{5} &E_{4} &0&E_{2} &E_{1} & 3 &E_{8} &E_{7} \\
     E_{7} &0 &E_{5} &E_{4} &0&E_{2} &E_{1} & 3 &E_{8} \\
     E_{8} & E_{7} &0 &E_{5} &E_{4} &0&E_{2} &E_{1} & 3
\end{array}
\right).
$$
   $$
   \hat{p}_{-2} = \left(
   \begin{array}{ccccccccc}
        1&0 &0 &3 E_{6} & 0&0 &3E_{3} &0 &0  \\
        0 & 1&0 &0 &3 E_{6} & 0&0 &3E_{3} &0 \\
        0 & 0& 1&0 &0 &3 E_{6} & 0&0 &3E_{3} \\
        3E_{3} &0 & 0& 1&0 &0 &3 E_{6} & 0&0 \\
        0& 3E_{3} &0 & 0& 1&0 &0 &3 E_{6} &0 \\
        0 & 0& 3E_{3} &0 & 0& 1&0 &0 &3 E_{6}\\
        3 E_{6} &0 & 0& 3E_{3} &0 & 0& 1&0 &0\\
        0 & 3 E_{6} &0 & 0& 3E_{3} &0 & 0& 1&0\\
        0 &0 & 3 E_{6} &0 & 0& 3E_{3} &0 & 0& 1\\
   \end{array}
   \right),
   $$
   $\hat{p} = \frac{1}{3} \hat{p}_{-1} + \frac{1}{9} \hat{p}_{-2}$,
$$
\hat{p} = \frac{1}{9}
\left(
\begin{array}{ccccccccc}
     4&E_{8} &E_{7} & 3 E_{6}& E_{5}& E_{4}&3 E_{3} &E_{2} & E_{1}\\
     E_{1} & 4 &E_{8} &E_{7} & 3 E_{6}& E_{5}& E_{4}&3 E_{3} &E_{2}\\
     E_{2} &E_{1} & 4 &E_{8} &E_{7} & 3 E_{6}& E_{5}& E_{4}&3 E_{3}\\
     3 E_{3}&E_{2} &E_{1} & 4 &E_{8} &E_{7} & 3 E_{6}& E_{5}& E_{4}\\
     E_{4} &3 E_{3}&E_{2} &E_{1} & 4 &E_{8} &E_{7} & 3 E_{6}& E_{5}\\
     E_{5} &E_{4} &3 E_{3}&E_{2} &E_{1} & 4 &E_{8} &E_{7} & 3 E_{6}\\
     3 E_{6} &E_{5} &E_{4} &3 E_{3}&E_{2} &E_{1} & 4 &E_{8} &E_{7} \\
     E_{7} &3 E_{6} &E_{5} &E_{4} &3 E_{3}&E_{2} &E_{1} & 4 &E_{8} \\
     E_{8} & E_{7} &3 E_{6} &E_{5} &E_{4} &3 E_{3}&E_{2} &E_{1} & 4
\end{array}
\right).
$$

\subsubsection{The case $n = 2$ and $N = 3^2 = 9$. Symmetric system}
$$\hat x=\hat x_0 + 3\hat x_1 = 
\text{diag}(4;3;2;1;0;-1;-2;-3;-4),$$
%=\left(
%    \begin{array}{ccccccccc}
%         4& 0 &0 &0 &0 &0 &0 &0 &0\\
%         0& 3 &0 &0 &0 &0 &0 &0 &0\\
%         0& 0 &2 &0 &0 &0 &0 &0 &0\\
%         0& 0 &0 &1 &0 &0 &0 &0 &0\\
%         0& 0 &0 &0 &0 &0 &0 &0 &0\\
%         0& 0 &0 &0 &0 &-1 &0 &0 &0\\
%         0& 0 &0 &0 &0 &0 &-2 &0 &0\\
%         0& 0 &0 &0 &0 &0 &0 &-3 &0\\
%         0& 0 &0 &0 &0 &0 &0 &0 &-4 \\
%    \end{array}
%    \right)
 %   \vspace{10truemm};$
$$\hat x_0=\text{diag}(1;0;-1;1;0;-1;1;0;-1),$$
%= \left(
%    \begin{array}{ccccccccc}
%         1& 0 &0 &0 &0 &0 &0 &0 &0\\
%         0& 0 &0 &0 &0 &0 &0 &0 &0\\
%         0& 0 &-1 &0 &0 &0 &0 &0 &0\\
%         0& 0 &0 &1 &0 &0 &0 &0 &0\\
%         0& 0 &0 &0 &0 &0 &0 &0 &0\\
%         0& 0 &0 &0 &0 &-1 &0 &0 &0\\
%         0& 0 &0 &0 &0 &0 &1 &0 &0\\
%         0& 0 &0 &0 &0 &0 &0 &0 &0\\
%         0& 0 &0 &0 &0 &0 &0 &0 &-1 \\
%    \end{array}
%    \right)$,
    
$$\hat x_1=\text{diag}(1;1;1;0;0;0;-1;-1;-1).$$
%= \left(
%    \begin{array}{ccccccccc}
%         1& 0 &0 &0 &0 &0 &0 &0 &0\\
%         0& 1 &0 &0 &0 &0 &0 &0 &0\\
%         0& 0 &1 &0 &0 &0 &0 &0 &0\\
%         0& 0 &0 &0 &0 &0 &0 &0 &0\\
%         0& 0 &0 &0 &0 &0 &0 &0 &0\\
%         0& 0 &0 &0 &0 &0 &0 &0 &0\\
%         0& 0 &0 &0 &0 &0 &-1 &0 &0\\
%         0& 0 &0 &0 &0 &0 &0 &-1 &0\\
%         0& 0 &0 &0 &0 &0 &0 &0 &-1\\
%    \end{array}
%    \right)$
 
Let's denote
$$
 G_{n} =  \frac{(-1)^{n+1}}{2\sqrt{3}\sin\left(\frac{\pi n}{9}\right)},
$$
 then
$$
\hat{p}_{-1} = \frac{1}{\sqrt{3}\I}\left(
\begin{array}{ccccccccc}
     0&G_{8} &G_{7} & 0& G_{5}& G_{4}&0 &G_{2} & G_{1}\\
     G_{1} & 0 &G_{8} &G_{7} & 0& G_{5}& G_{4}&0 &G_{2}\\
     G_{2} &G_{1} & 0 &G_{8} &G_{7} & 0& G_{5}& G_{4}&0\\
     0&G_{2} &G_{1} & 0 &G_{8} &G_{7} & 0& G_{5}& G_{4}\\
     G_{4} &0&G_{2} &G_{1} & 0 &G_{8} &G_{7} & 0& G_{5}\\
     G_{5} &G_{4} &0&G_{2} &G_{1} & 0 &G_{8} &G_{7} & 0\\
     0 &G_{5} &G_{4} &0&G_{2} &G_{1} & 0 &G_{8} &G_{7} \\
     G_{7} &0 &G_{5} &G_{4} &0&G_{2} &G_{1} & 0 &G_{8} \\
     G_{8} & G_{7} &0 &G_{5} &G_{4} &0&G_{2} &G_{1} & 0
\end{array}
\right),
$$
$$
   \hat{p}_{-2} = -\frac{1}{\sqrt{3}\I} \left(
   \begin{array}{ccccccccc}
        0&0 &0 &1 & 0&0 &-1  &0 &0  \\
        0 & 0&0 &0 &1 & 0&0 &-1  &0 \\
        0 & 0& 0&0 &0 &1 & 0&0 &-1  \\
        -1 &0 & 0& 0&0 &0 &1 & 0&0 \\
        0& -1  &0 & 0& 0&0 &0 &1 &0 \\
        0 & 0& -1  &0 & 0& 0&0 &0 &1\\
        1 &0 & 0& -1  &0 & 0& 0&0 &0\\
        0 & 1 &0 & 0& -1  &0 & 0& 0&0\\
        0 &0 & 1 &0 & 0& -1  &0 & 0& 0\\
   \end{array}
   \right);
   $$
   $\hat{p} = \frac{1}{3} \hat{p}_{-1} + \frac{1}{9} \hat{p}_{-2}$,
$$
\hat{p} = \frac{1}{9\sqrt{3}\I}\left(
\begin{array}{ccccccccc}
     0&3G_{8} &3G_{7} & -1& 3G_{5}& 3G_{4}&1&3G_{2} & 3G_{1}\\
     3G_{1} & 0 &3G_{8} &3G_{7} & -1& 3G_{5}& 3G_{4}&1 &3G_{2}\\
     3G_{2} &3G_{1} & 0 &3G_{8} &3G_{7} & -1& 3G_{5}& 3G_{4}1\\
     1&3G_{2} &3G_{1} & 0 &3G_{8} &3G_{7} & -1& 3G_{5}& 3G_{4}\\
     3G_{4} &1&3G_{2} &3G_{1} & 0 &3G_{8} &3G_{7} & -1& 3G_{5}\\
     3G_{5} &3G_{4} &1&3G_{2} &3G_{1} & 0 &3G_{8} &3G_{7} & -1\\
     -1 &3G_{5} &3G_{4} &1&3G_{2} &3G_{1} & 0 &3G_{8} &3G_{7} \\
     3G_{7} &-1 &3G_{5} &3G_{4} &1&3G_{2} &3G_{1} & 0 &3G_{8} \\
     3G_{8} & 3G_{7} &-1 &3G_{5} &3G_{4} &1&3G_{2} &3G_{1} & 0
\end{array}
\right).
$$

\end{document}